\documentclass[notitlepage]{iopart}
\usepackage
[
        a4paper,
        left=3cm,
        right=3cm,
        top=3cm,
        bottom=4cm,
]{geometry}
 
\expandafter\let\csname equation*\endcsname\relax
\expandafter\let\csname endequation*\endcsname\relax
\expandafter\let\csname eqnarray*\endcsname\relax
\expandafter\let\csname endeqnarray*\endcsname\relax
\usepackage{amsmath}
\usepackage{undertilde}
\usepackage{amssymb}
\usepackage{graphicx}

\begin{document}
\title{Gravitational properties of light - The gravitational field of a laser pulse}
\author{Dennis R\"atzel, Martin Wilkens, Ralf Menzel}
\address{University of Potsdam, Institute for Physics and Astronomy\\
Karl-Liebknecht-Str. 24/25, 14476 Potsdam, Germany}
\ead{raetzel@uni-potsdam.de}
\begin{abstract}
The gravitational field of a laser pulse of finite lifetime, is investigated in the framework of linearized gravity. Although the effects are very small, they may be of fundamental physical interest. It is shown that the gravitational field of a linearly polarized light pulse is modulated as the norm of the corresponding electric field strength, while no modulations arise for circular polarization. In general, the gravitational field is independent of the polarization direction.
It is shown that all physical effects are confined to spherical shells expanding with the speed of light, and that these shells are associated with the emission and absorption of the pulse. 
Nearby test particles at rest are attracted towards the pulse trajectory by the gravitational field due to the emission of the pulse, and they are repelled from the pulse trajectory by the gravitational field due to its absorption. Examples are given for the size of the attractive effect. It is recovered that massless test particles do not experience any physical effect if they are co-propagating with the pulse, and that the acceleration of massless test particles counter-propagating with respect to the pulse is four times stronger than for massive particles at rest. The similarities between the gravitational effect of a laser pulse and Newtonian gravity in two dimensions are pointed out. The spacetime curvature close to the pulse is compared to that induced by gravitational waves from astronomical sources.\\

\end{abstract}



\section{Introduction}
\label{sec:introduction}

A pulse of light carries energy and momentum, and it is deflected by the gravitational field of a massive body. Then, according to Newton's ``actio equals reactio'', it is also source of a gravitational field of its own.
The gravitational field of a laser pulse may hardly be detectable under laboratory conditions, but it comes with some peculiar features which are of general physical interest. Utilizing linearized gravity, in \cite{Tolman1931} by Tolman, Ehrenfest and Podolsky it was established that the gravitational field of a cylindrical pulse of unpolarized light, of finite lifetime, for which diffraction can be neglected does not affect a parallel test beam if the test beam is co-propagating, but bends it, if counter-propagating. Stated differently, a freely propagating light pulse would not be affected by its own gravitational field, which is in sharp contrast to a beam of massive particles.

In a series of subsequent investigations, the gravitational field of light has been determined within the framework of the full set of the nonlinear Einstein equations in which light is represented as a null-fluid of massless particles \cite{Bonnor1969}, from the Lorentz-boosted Schwarzschild-metric of a point mass in the limit $v\rightarrow c$, $m\rightarrow 0$ \cite{Aichelburg1971}, and even some exact plane wave solutions of the coupled Maxwell-Einstein theory \cite{vanHolten2011}. It is now well established that the gravitational field of light is twice that of a material source of the same energy-mass density, that a pulse of light on an infinite straight path is accompanied by a co-propagating plane fronted gravitational wave, and that two such pulses would never interact if propagating on parallel tracks in the same direction \footnote{The case of counter-propagating light beams has been settled  in \cite{Kramer1998} by Kramer.}. In \cite{Scully1979} by Scully, it was shown that the interaction between pulses running slower than the speed of light - e.g. in a wave guide - is non-zero, however. In \cite{Aichelburg1971} by Aichelburg and Sexl, it is pointed to the inadequacy of the standard Greens function method for the solution to linearized gravity with sources on strict null-paths, i.e for light pulses  in vacuum on infinite straight paths. Following early studies \cite{Hegarty1969}, the critical role of the pulse emission and absorption for the gravitational effects of light was emphasized in \cite{Voronov1973} by Voronov and Kobzarev. The recent publication \cite{Bonnor2009} by Bonnor comes to the conclusion that even within the full set of Einstein equations, the gravitational field of a laser pulse on an infinite straight path cannot arise from the retarded potential generated by the pulse, but should be ascribed to the process of its emission \footnote{In \cite{Azzurli2014} by Azzurli and Lechner, similar conclusions are drawn for the electromagnetic field generated by massless charged particles in the context of classical Maxwell theory.}.

In this paper, we derive the gravitational field which comes with a laser pulse, using the framework of linearized gravity. In accordance with the established model, the pulse is represented as a ``needle of null stuff'', in our case consisting of coherently polarized electromagnetic radiation.
Our model aims to catch the essential ingredient of a laser pulse, which are (1) its localizability, (2) its masslessness, and (3) its polarization degrees of freedom.
The process of emission and absorption of the pulse is also included. This accounts for the very nature of the electromagnetic field as the mediator of the electromagnetic interaction between material bodies and at the same time avoids the above mentioned calamities which come with massless excitations on unbound trajectories.

The paper is organized as follows. In Section \ref{sec:pulse}, we review linearized gravity and derive the metric perturbation which is caused by a laser pulse. In sections \ref{sec:geodesics} to \ref{sec:defltow}, we investigate the gravitational effect of the pulse on a test particle moving along a geodesic in the gravitational field of the laser pulse. In section \ref{sec:conclusions}, we discuss the gravitational effect of laser pulses in a broader context and show possibilities for further investigations.

\section{Laser pulse metric perturbation}
\label{sec:pulse}

In this section, we derive the gravitational field of a pulse of laser light. We assume the power of the pulse to be small such that the metric tensor $g_{\mu\nu}$ differs but slightly from the Minkowski metric in free space which takes values $\eta_{\mu\nu}=\mathrm{diag}(-1,1,1,1)$ for appropriately chosen coordinates $(ct,x,y,z)$. Setting
\begin{equation}\label{eq:metricsplit}
	g_{\mu\nu} = \eta_{\mu\nu} + h_{\mu\nu}\,,
\end{equation}
where $h_{\mu\nu}$ is a small perturbation, $|h_{\mu\nu}|\ll 1$, and imposing the gauge condition
\begin{equation}\label{eq:lorentzgauge}
	\partial^\mu \left(h_{\mu\nu} - \frac{1}{2}\eta_{\mu\nu} {h_\alpha}^\alpha\right)=0\,,
\end{equation}
where ${h_\alpha}^\alpha=h_{\alpha\beta}\eta^{\alpha\beta}$, the Einstein field equations assume the well known form of linearized gravity
\begin{equation}\label{eq:linearizedeinstein}
	\left[\frac{\partial^2}{\partial x^2}
	+ \frac{\partial^2}{\partial y^2}
	+ \frac{\partial^2}{\partial z^2}
	- \frac{1}{c^2}\frac{\partial^2}{\partial t^2}
	\right] h_{\mu\nu} = \frac{16\pi G}{c^4}
	\left(T_{\mu\nu}-\frac{1}{2}\eta_{\mu\nu} {T_\alpha}^\alpha\right)\,,
\end{equation}
where $T_{\mu\nu}$ is the energy-momentum tensor of the pulse, including pulse emitter and absorber, $G$ is Newton's gravitational constant, $c$ is the speed of light and ${T_\alpha}^\alpha=T_{\alpha\beta}\eta^{\alpha\beta}$. We assume energy-momentum conservation,
\begin{equation}
	\partial^\nu T_{\mu\nu} = 0\,,
\end{equation}
such that consistency with the gauge-condition (\ref{eq:lorentzgauge}) is guaranteed. 

\begin{figure}[h]
\hspace{2.2cm}
\includegraphics[width=8cm,angle=0]{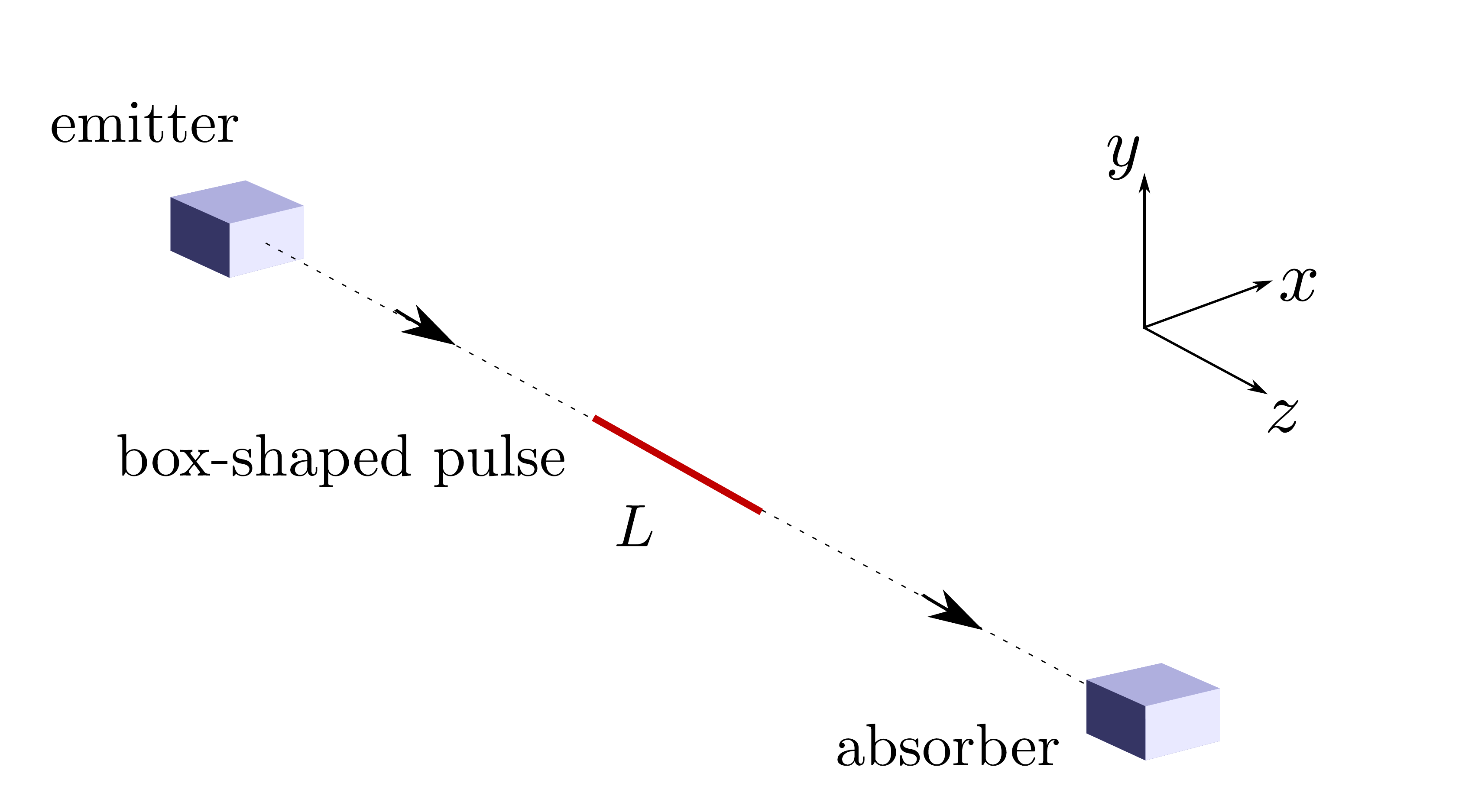}
\caption{The laser pulse is modeled as a pulse of electromagnetic radiation of length $L$, traveling from emitter to absorber over a distance $D$ along the $z$-axis. The extension of the pulse in the transverse $x/y$-directions is assumed to be negligible in comparison to its length.
 \label{fig:pulse}}
\end{figure}
The laser pulse is modeled as a pulse of electromagnetic radiation, traveling from emitter to absorber over a distance $D$ along the $z$-axis, with finite extension (pulse length) $L$ in the direction of propagation, but negligible extension $\Delta(z)$ in the transverse $x/y$-directions, $\Delta(z)\ll L$ (see figure \ref{fig:pulse}). All measures refer to a laboratory frame where the emitting system is at rest before emission of the pulse. Introducing the formal decomposition
\begin{equation}\label{eq:splitpea}
	T_{\mu\nu} = T^\mathrm{p}_{\mu\nu} + T^\mathrm{e}_{\mu\nu} + T^\mathrm{a}_{\mu\nu}
\end{equation}
with $T^\mathrm{e}_{\mu\nu}$ and $T^\mathrm{a}_{\mu\nu}$ the energy-momentum of the physical systems which are involved in the emission and absorption, respectively, and the pure pulse contribution $T_{\mu\nu}^\mathrm{p}$ which is non-zero only during the pulse lifetime which is of order $c^{-1}(L+D)$. More important, for all times preceding the pulse emission, the energy-momentum tensor support is confined to a region which is spatially finite, and thus we may safely invoke the standard Green's function method to solve equation (\ref{eq:linearizedeinstein}). Being primarily interested in the metric perturbations which are causally connected to the source, the corresponding solution of (\ref{eq:linearizedeinstein}) reads
\begin{equation}\label{eq:retarded}
	h_{\mu\nu}(x,y,z,t)
	=
	\frac{4 G}{c^4} \int  \frac{
	\left(T_{\mu\nu}-\frac{1}{2}\eta_{\mu\nu}{T_\alpha}^\alpha\right)(x',y',z',t_\mathrm{ret})}{\sqrt{(x-x')^2+(y-y')^2+(z-z')^2}} dx'dy'dz'\,,
\end{equation}
with $t_\mathrm{ret}$ the retarded time, $t_\mathrm{ret}=t-\sqrt{(x-x')^2+(y-y')^2+(z-z')^2}/c$.

\begin{figure}[h]
\hspace{2cm}
\includegraphics[width=8cm,angle=0]{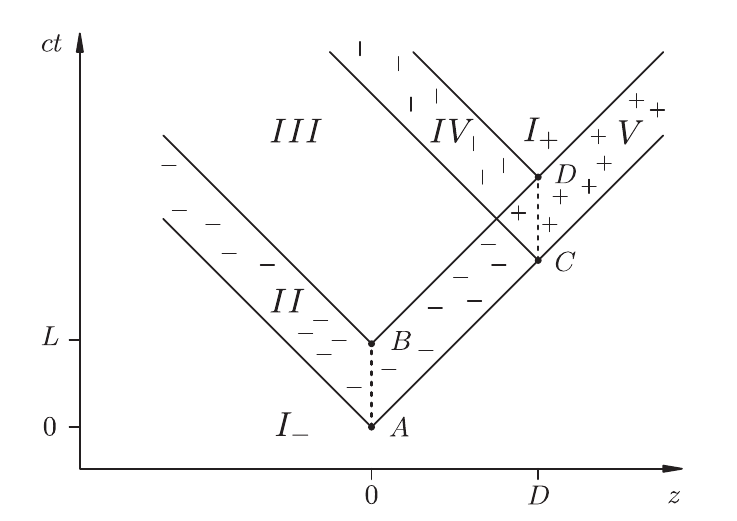}
\caption{ The pulse defines a world sheet restricted to the $t$-$z$-plane. The world sheet is spanned between the points A, B, C and D which correspond, respectively, to the start of the pulse emission, the end of the emission, the start of the absorption and the end of the absorption. The future directed light cones of A, B, C, D define the spacetime regions $I_-$-$V$ with qualitatively different metric perturbations.
 \label{fig:Regions}}
\end{figure}

Our pulse model is further specified by ``boxing'' the energy-momentum tensor of a monochromatic plane wave with an appropriate envelope function $A \chi(z,t) \delta(x)\delta(y)$, where $A$ is an effective area of the pulse transverse extend, and $\chi(z,t)$ a characteristic function (normalized $\chi^2=\chi$) which -- for any given time $t$ -- encodes the momentary extension and location of the laser pulse on the $z$-axis. With the emitter and absorber placed at fixed positions $z=0$ and $z=D$, respectively, and fixing the time coordinate such that the laser pulse appears at $z=0$ at time $t=0$ and thus the back of our pulse -- which in free flight is of length $L$ -- leaves the emitter at $z=0$ at a later time $T=L/c$, the entire pulse defines a world sheet, restricted to the  $t$-$z$-plane -- see figure~\ref{fig:Regions}.

Recall that, for electromagnetic plane waves propagating in the $z$-direction, the corresponding energy-momentum tensor depends only on the combination $ct-z$, and the only non-vanishing components are given by $T_{00}= T_{zz} = - T_{0z} = -T_{z0} =u$, with $u=\frac{\varepsilon_0}{2}\vec{E}^2 + \frac{1}{2\mu_0}\vec{B}^2$ the energy density of the electromagnetic field, where the index $0$ corresponds to $ct$. Accordingly, the only non-vanishing components of the pulse contribution to the metric perturbation are 
\begin{equation}\label{eq:hcomponents}
 h_{00}^\mathrm{p} = h_{zz}^\mathrm{p} = -h_{0z}^\mathrm{p}=-h_{z0}^\mathrm{p}=h^\mathrm{p}\,,
\end{equation}
where $h^\mathrm{p}$ can be read off from (\ref{eq:retarded})
\begin{equation}\label{eq:retardedzprime}
	h^\mathrm{p}(x,y,z,t) = \frac{4GA}{c^4} \int_a^b
	\frac{u(ct_\mathrm{ret}(z')-z')}{\sqrt{\rho^2+(z'-z)^2}} dz'
\end{equation}
with $\rho=\sqrt{x^2+y^2}$ the observer's distance to the axis of pulse propagation, and $t_\mathrm{ret}(z')=t-\sqrt{\rho^2+(z-z')^2}/c$.  We see in equation (\ref{eq:retardedzprime}) that the condition of $h_{\mu\nu}$ being small translates to the power of the laser pulse being small.

\begin{figure}[h]
\hspace{2.3cm}
\includegraphics[width=8cm,angle=0]{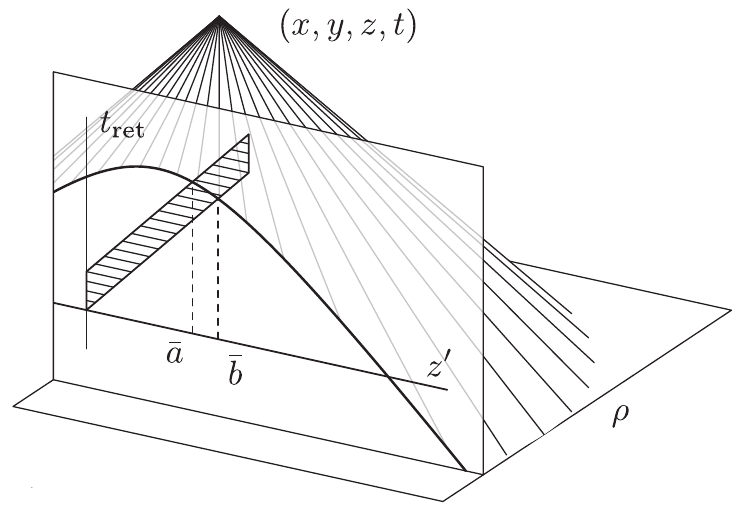}
\caption{ The integral boundaries $a,b$ are determined by the intersection of the pulse world sheet boundaries with the past light cone of the observation event $x,y,z,t$.
 \label{fig:3DCone}}
\end{figure}
The integral boundaries $a,b$ are determined by the intersection of the pulse world sheet boundaries with the past light cone of the observation event $x,y,z,t$ (see figure \ref{fig:3DCone}). They are most conveniently described in terms of the auxiliary space-time functions
\begin{eqnarray}
	\bar{a} & = &  z + \frac{(ct-L-z)^2 - \rho^2}{2(ct-L-z)}\,, 
	\\
	\bar{b} & = & z + \frac{(ct-z)^2 - \rho^2}{2(ct-z)}\,,
\end{eqnarray}
which are the solutions of $t_\mathrm{ret}(z')=(z'+L)/c$ and $t_\mathrm{ret}(z')=z'/c$, respectively. Depending on the coordinates of the observation event we then have $[a,b]=\emptyset$ if both $\bar{a}$ and $\bar{b}$ are smaller $0$ (Region $I_-$) or both are larger $D$ (Region $I_+$), and 
\begin{equation}\label{eq:regions}
	[a,b] = \left\{
	\begin{array}{ccrl}
		\left[0,\bar{b}\right] & \,, & \bar{a}<0<\bar{b}<D & \mbox{\ (Region II)}
		\\
		\left[\bar{a},\bar{b}\right] & \,, & 0<\bar{a}<\bar{b}<D & \mbox{\ (Region III)}
		\\
		\left[\bar{a},D\right] & \,, & 0<\bar{a} <D<\bar{b} & \mbox{\ (Region IV)}
		\\
		\left[0,D\right] &\,, & \bar{a}<0<D<\bar{b} & \mbox{\ (Region V)}
	\end{array}
	\right.\,.
\end{equation}
The meaning of the various regions derives from the causal relationship of the observation event to the laser pulse (see figure \ref{fig:Regions}). Region $I_-$ and $I_+$ are completely disconnected from the system. In space-time region II, details of the pulse emission are witnessed, while in region IV it is the pulse absorption and in region $V$ it is both, emission and absorption. Region III is completely disconnected from pulse emission and absorption, but only experiences the passage of the pulse. Note that in any case, due to the invariance under rotations around the $z$-axis, the metric perturbation only depends on $\rho$, $z$ and $t$, but not on the azimuthal angle $\varphi$.

To further elucidate the coordinate dependence, we utilize a variable substitution $z'\mapsto \zeta$,
\begin{equation}\label{eq:zeta}
	\zeta(z') = (z'-z) + \sqrt{\rho^2 + (z'-z)^2}\,,
\end{equation}
such that
\begin{equation}\label{eq:hint}
	h^\mathrm{p}(x,y,z,t) = \frac{4GA}{c^4} \int_{\zeta(a)}^{\zeta(b)}
	\frac{u(z-ct+\zeta)}{\zeta} d\zeta\,.
\end{equation}
Noteworthy, in Region III,  we have $\zeta(a)\equiv\zeta(\bar{a})= ct-z-L$ and $\zeta(b)\equiv\zeta(\bar{b})=ct-z$. Hence in this region, where the observation event is causally disconnected from both the emission and absorption, the metric perturbation depends exclusively on the light cone coordinate $ct-z$. This dependence, however, has no physical effects, as demonstrated in the next section. 

In the following, we shall frequently refer to two cases -- circular polarization and linear polarization of the laser pulse. In the case of circular polarization the energy density is constant, i.e $u(z-ct)=u_0$, but in case of linear polarization the energy density is sinusoidally modulated, $u(z-ct)= 2u_0\sin^2(\omega( t - z/c ) + \varphi)$. Noteworthy, for both cases, linear polarization and circular polarization, the remaining integral (\ref{eq:hint}) can be performed in closed form. In case of circular polarization, for example, the contribution of the pulse to the metric perturbation is given by the simple expression
\begin{equation}\label{eq:hintcirc}
	h^\mathrm{p}(x,y,z,t) = \kappa \ln\left(\frac{\zeta(b)}{\zeta(a)}\right)\,,
\end{equation}
with constant $\kappa=4GAu_0/c^4$. In the case of linear polarization, the right hand side is replaced by linear combinations of integral sine and cosine functions -- see Appendix B for details. 

\begin{figure}[h]
\hspace{2.2cm}
\includegraphics[width=10cm,angle=0]{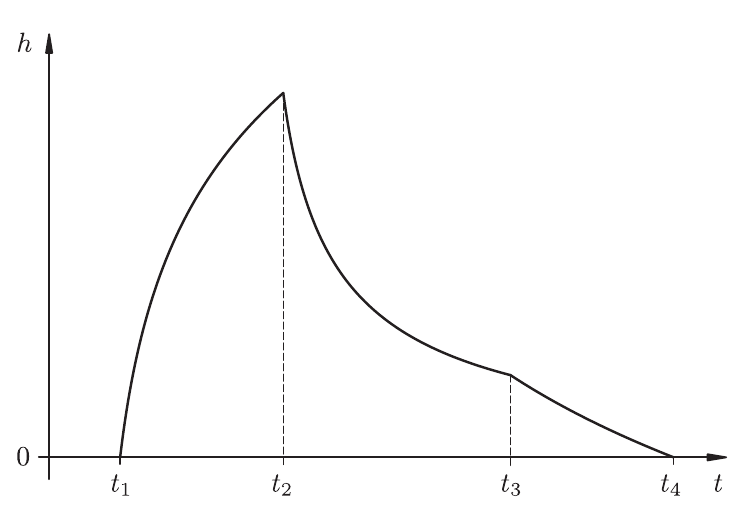}
\caption{The temporal evolution of $h^\mathrm{p}$ for an observer at fixed position $(x,y,z)$ passing consecutively through Region $I_-$ ($t<t_1$), Region $II$ ($t_1<t<t_2$), Region $III$ ($t_2<t<t_3$), Region $IV$ ($t_3<t<t_4$) and Region $I_+$ ($t_4<t$).
 \label{fig:h_of_t}}
\end{figure}

For an observer at fixed position $(x,y,z)$, the temporal evolution of $h^\mathrm{p}$ is displayed in figure \ref{fig:h_of_t}.
For early times $t<t_1= \frac{1}{c}\sqrt{\rho^2+z^2}$, the metric perturbation is zero. For times $t_1<t<t_2=t_1+L/c$, the observer bears witness of the photon emission, and the metric perturbation rises from zero to its maximal value. In the subsequent range $t_2<t<t_3=D/c+\sqrt{\rho^2+(z-D)^2}/c$, the metric perturbation decays slowly $h^\mathrm{p}=\kappa \ln\left(1+\frac{L}{ct-z-L}\right)$. Beginning with $t_3$ and ending with $t_4=t_3+L/c$, the observer experiences the photon absorption, and the metric perturbation decays back to the zero value, which it maintains for all times $t>t_4$.

\begin{figure}[h]
\hspace{2.1cm}
\includegraphics[width=6cm,angle=0]{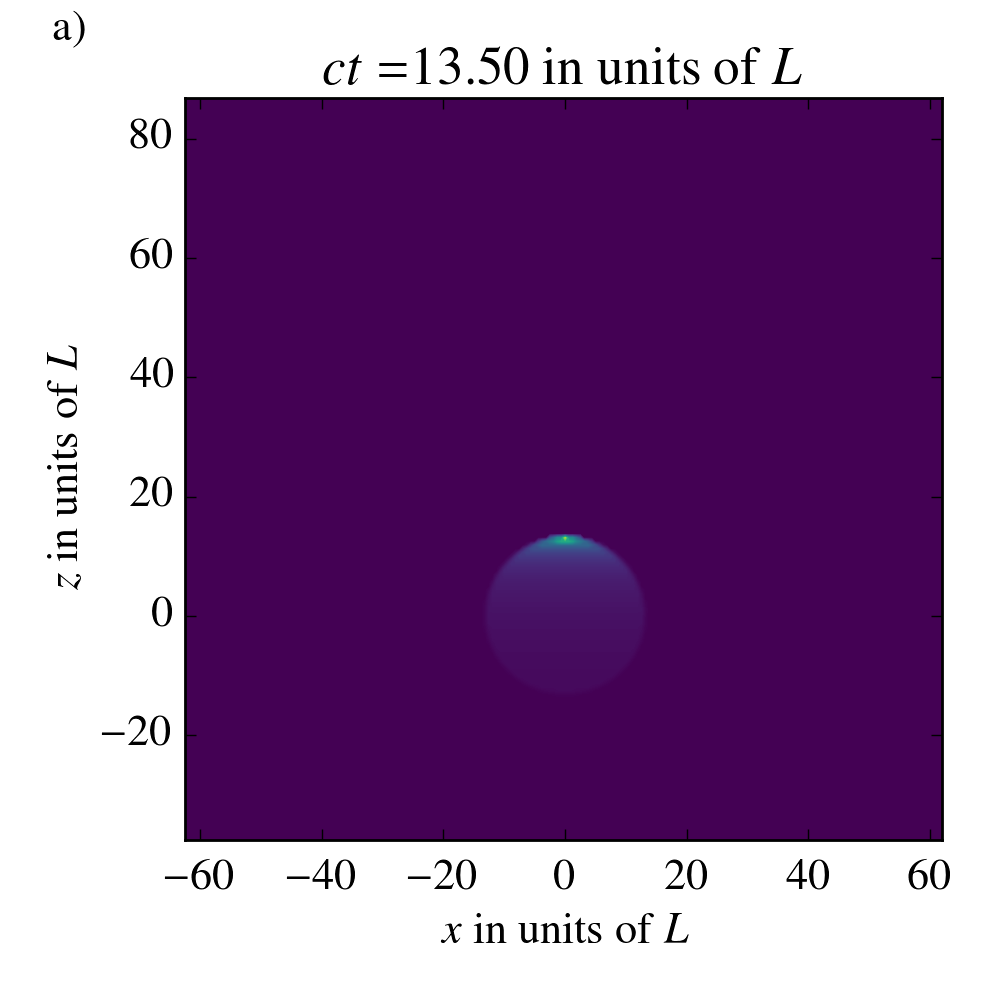}
\includegraphics[width=7.2cm,angle=0]{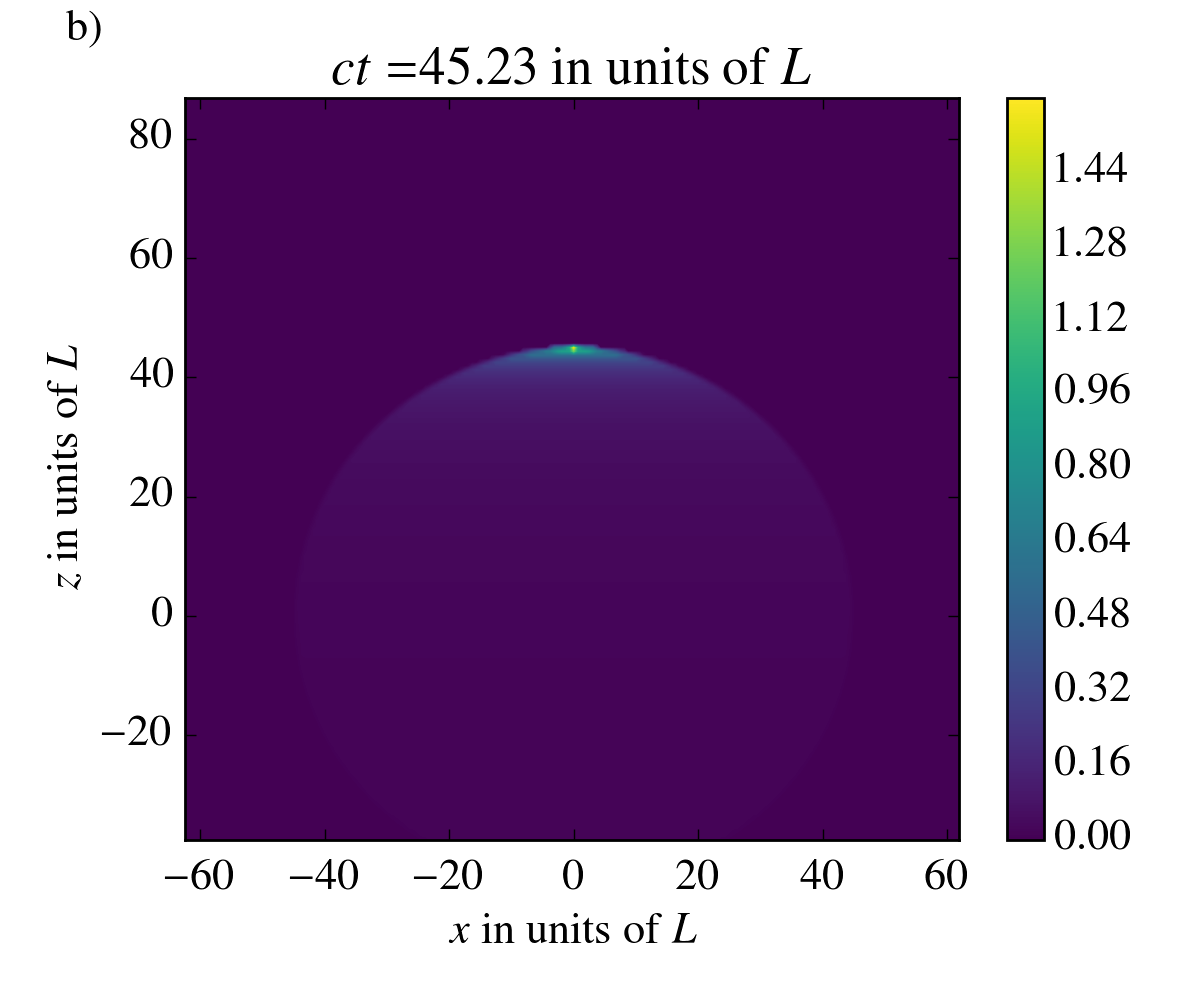}

\hspace{2.1cm}
\includegraphics[width=6cm,angle=0]{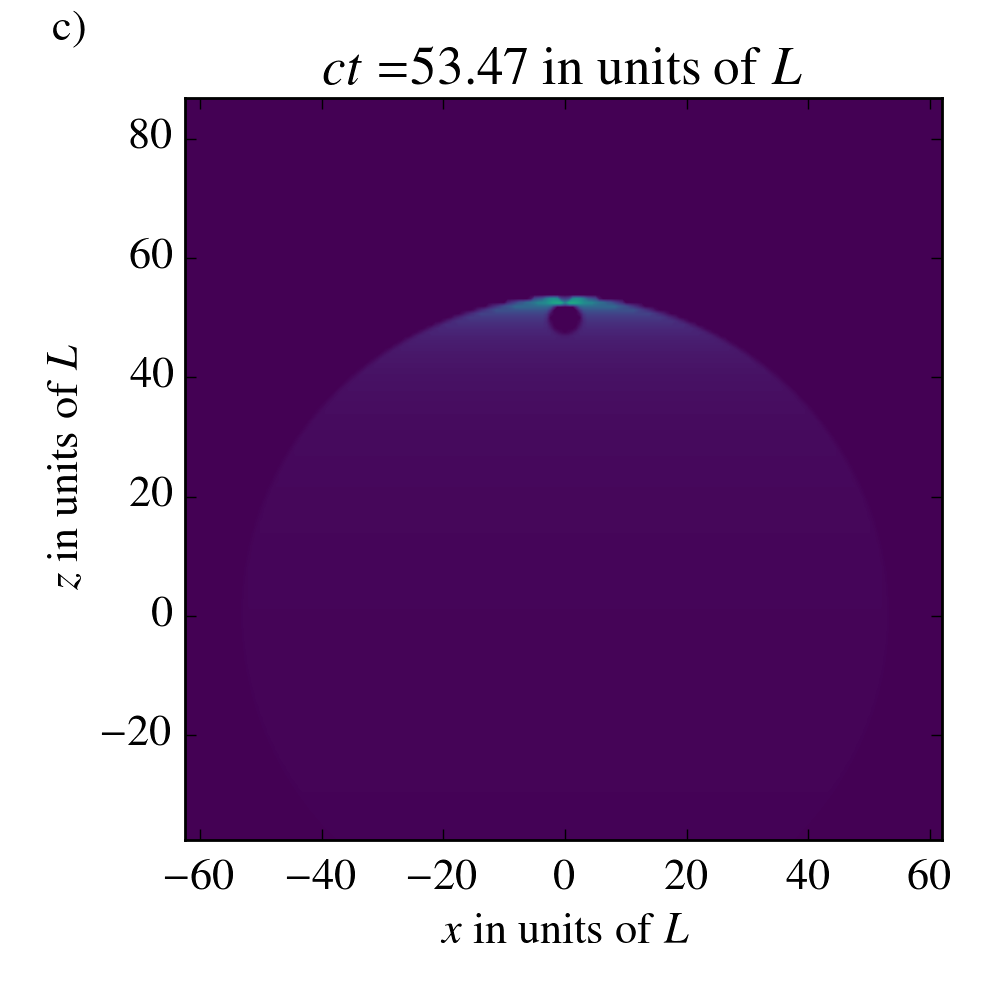}
\includegraphics[width=7.2cm,angle=0]{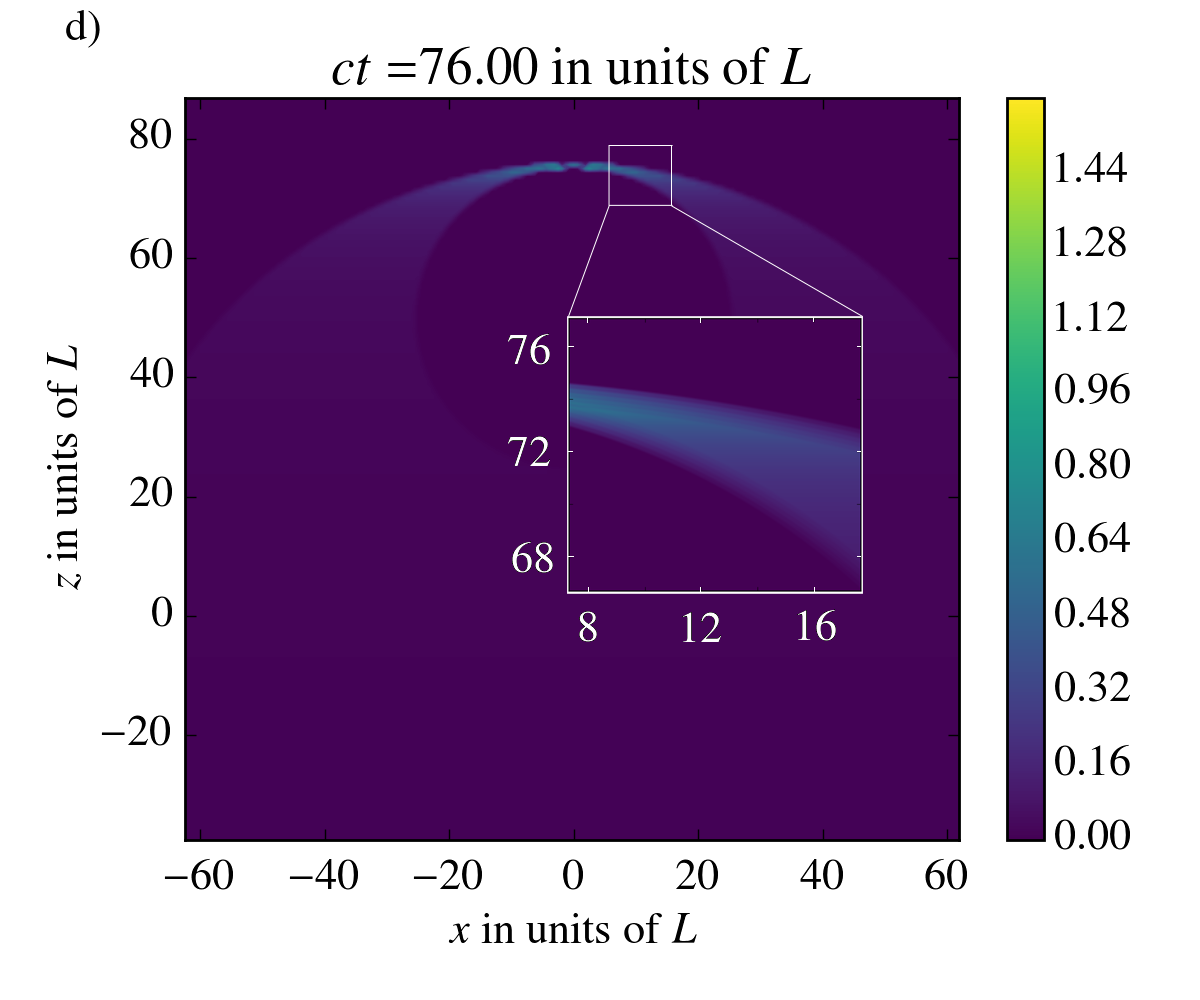}
\caption{\label{fig:hval} The plots show the metric perturbation $h^\mathrm{p}=h_{00}^\mathrm{p} = h_{zz}^\mathrm{p} = -h_{0z}^\mathrm{p}=-h_{z0}^\mathrm{p}$ for a pulse of length $L$ in the coordinates $(ct,x,y,z)$ in the $(x,y)$-plane for different times $t$. $h^\mathrm{p}$ is normalized to units of $\kappa$ and then the logarithm of the logarithm is taken. In a), we see the effect of the pulse on the metric expanding from the point of the emission of the pulse at $z=0$. In b), the field has already expanded. In c), the pulse has been annihilated at $x=D$ and the sphere expands that contains the information about the absorption. In d), this sphere has further expanded and both fronts approach the form of a plane fronted wave.}
\end{figure}
In figure \ref{fig:hval}, $h^\mathrm{p}$ is plotted in the $x$-$z$-plane for different times after the emission of the pulse. The corresponding spatial sections of the regions $I_-$-$V$ can be identified, and the influence of the pulse emission/absorption can be seen to propagate spherically with the speed of light from the point of creation/absorption. For distances to the trajectory of the pulse much smaller than the distance from the observer to the emitter, the gravitational field approaches the form of a plane fronted parallel propagating wave (pp-wave) \cite{Bonnor1969}.

It is also interesting to notice that after the absorption of the pulse, a non-zero metric perturbation remains (see figure \ref{fig:hval} d). However, we find that the physical effect decays in the limit $t\rightarrow \infty$ staying at a fixed distance to the $z$-axis since the front of emission and the front of absorption become planes and coincide eventually in this limit. We will see later that the gravitational effect decays, basically, with the inverse of the distance to the trajectory of the pulse. We conclude that for long times after the absorption no gravitational effect remains.

\begin{figure}[h]
\hspace{2.1cm}
\includegraphics[width=6.5cm,angle=0]{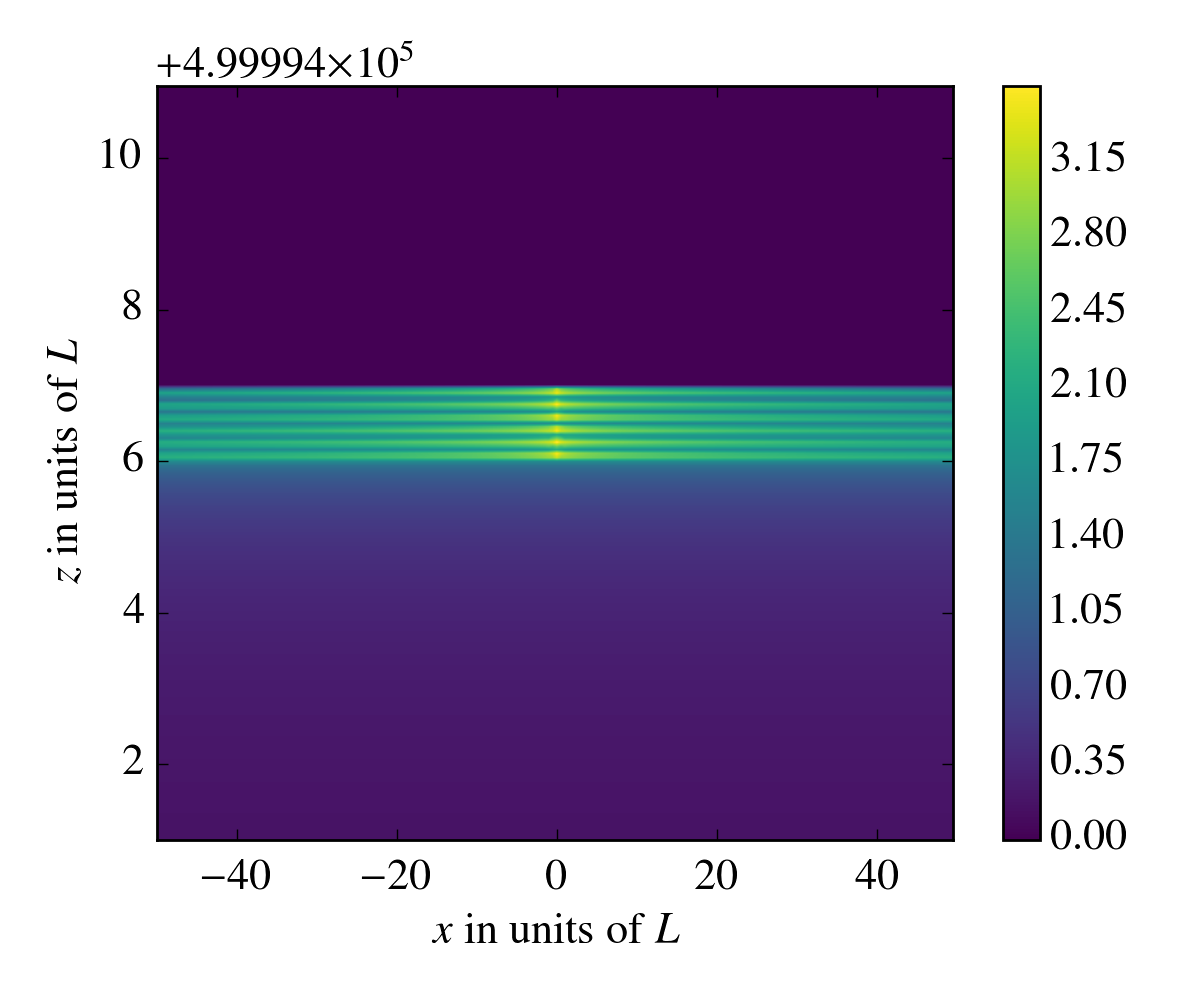}
\includegraphics[width=6.5cm,angle=0]{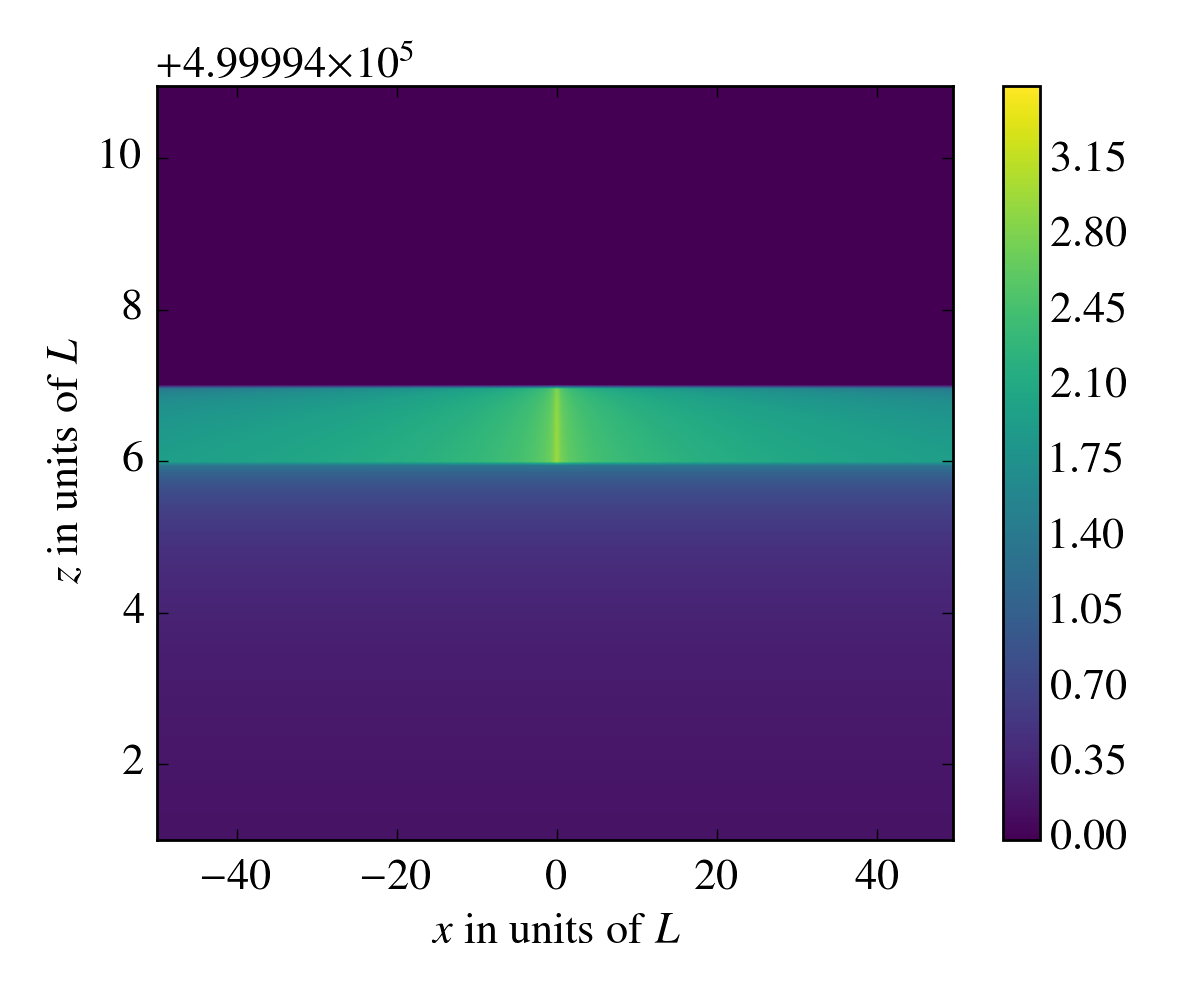}
\caption{\label{fig:photon_long_lifetime} These plots show the double logarithm of the metric perturbation $h^\mathrm{p}=h_{00}^\mathrm{p} = h_{zz}^\mathrm{p} = -h_{0z}^\mathrm{p}=-h_{z0}^\mathrm{p}$ for a linearly polarized pulse of length $L$ and central wavelength $\lambda=\frac{2\pi c}{\omega}=\frac{2}{3}L$ in the $x$-$y$-plane at $t=50000L/c$, after its emission at $z=0$. $h^\mathrm{p}$ is normalized to units of $\kappa=4GAu_0/c^4$ and then the logarithm of the logarithm is taken. The metric perturbation can be interpreted as the potential to the gravitational field. The front stemming from the emission of the pulse is seen between $z=6L+499994L$ and $z=7L+499994L$. It shows oscillations with wavelength $\lambda/2$ and approaches the form of a plane fronted wave. The right plot shows the same situation for circularly polarized light where no modulations appear.}
\end{figure} 
In figure \ref{fig:photon_long_lifetime}, the metric perturbation is plotted for linearly and circularly polarized light and $\rho/r\ll 1$. The metric perturbation is modulated as the norm of the electric field strength ($|E|^2$) for linearly polarized light. For circularly polarized light no modulation arises.

Yet, equations (\ref{eq:hcomponents})--(\ref{eq:hint}) only describe the contribution of the freely traveling pulse to the metric perturbation. For a rigorous account of the gravitational effect, also the contribution of the emitter and absorber must be taken into account. Indeed, by itself, $T^\mathrm{p}_{\mu\nu}$ violates energy momentum conservation, $\partial^\nu T^\mathrm{p}_{\mu\nu}\neq 0$ locally along the world lines of the emitting and absorbing phase, and -- concomitantly -- the Lorentz gauge condition (\ref{eq:lorentzgauge}) is violated.

A massive object, upon emission of a laser pulse of energy $E$, suffers a loss of rest mass $M\rightarrow M'=M\sqrt{1-2E/(Mc^2)}$, and at the same time acquires recoil momentum $p\rightarrow p'=p-E/c$. The reduction of rest mass implies a reduction of the gravitational field of the emitter, and this change travels with the same speed of light as does the emitted pulse. Yet in contrast to the gravitational field due to the pulse, which is stronger the closer we are to the pulse
(represented in its $z-r$-dependence (see equations (\ref{eq:zeta}) and (\ref{eq:hint}) for $a=0$ or $b=D$))
, the change of the emitter's gravitational field falls off with the spatial distance to the emitter which is approximately $r$ if we assume $Mc^2\gg E$. 
Hence, at points sufficiently close to the axis of the pulse propagation, i.e. $\rho/r\ll 1$, the passing gravitational field of the pulse outweighs the concomitant change in the emitters gravitational potential, such that the latter effect may safely be ignored. The same argument applies for the absorption.

In the next sections, we will investigate the behavior of particles in the gravitational field of the pulse.

\section{Test particles in the gravitational field of the laser pulse}
\label{sec:geodesics}

In this section, we investigate the behavior of freely falling test particles in the gravitational field of a laser pulse. Recall that in general relativity, the space-time geometry is coded in the line element $ds^2=g_{\mu\nu}dx^\mu dx^\nu$, and the world line $\gamma^\mu(\lambda)$ of free test particles is governed by the geodesic equations 
\begin{equation}\label{eq:geodesiceq}
\ddot{\gamma}^\mu = -\Gamma^\mu_{\rho\sigma}\dot{\gamma}^\rho\dot{\gamma}^\sigma
\end{equation}
with $g_{\mu\nu}\dot{\gamma}^\mu\dot{\gamma}^\nu=-1$ for massive test particles, and $g_{\mu\nu}\dot{\gamma}^\mu\dot{\gamma}^\nu = 0$ for massless test particles. Here, the dot indicates the derivative with respect to the curve parameter, which is proper time $\tau$ for the time-like geodesics of massive particles. For the null-geodesics of massless particles, we will use coordinate time $t$. Finally, $\Gamma^\mu_{\rho\sigma}$ are the Christoffel symbols, $\Gamma^{\mu}_{\rho\sigma} = \frac{1}{2} g^{\mu\nu}\left(\partial_\rho g_{\sigma\nu} + \partial_{\sigma} g_{\nu\rho} -\partial_\nu g_{\rho\sigma}\right)$.

For the metric perturbation of our light pulse, the line element reads
\footnote{In light-cone coordinates $u:=ct-z$, $v:=ct+z$ the line element reads $ds^2 = -du dv + h^\mathrm{p} du^2 + dx^2 + dy^2$.}
\begin{equation}
   ds^2 = -(1-h^\mathrm{p}) c^2 dt^2 + (1+h^\mathrm{p}) dz^2 - 2h^\mathrm{p} cdt dz + dx^2 + dy^2\,.
\end{equation}
and the geodesic equations are given as
\begin{eqnarray}\label{eq:geodesicdevexpl}
 \nonumber  \ddot{\gamma^z} & = & \textstyle{\frac{1}{2}} \partial_z h^\mathrm{p}  (\dot{\gamma}^u)^2
   + \textstyle{\frac{d}{d\lambda}}\left(h^\mathrm{p}\dot{\gamma}^u\right)
   \\
   \ddot{\gamma^x} & = & \textstyle{\frac{1}{2}} \partial_x h^\mathrm{p}  (\dot{\gamma}^u)^2
   \\
 \nonumber  \ddot{\gamma^y} & = & \textstyle{\frac{1}{2}} \partial_y h^\mathrm{p} (\dot{\gamma}^u)^2
\end{eqnarray}
with subsidiary condition
\begin{equation}\label{eq:subsed}
   -\dot{\gamma}^u\dot{\gamma}^v + h^\mathrm{p}(\dot{\gamma}^u)^2 + (\dot{\gamma}^x)^2 + (\dot{\gamma}^y)^2 =
   \left\{\begin{array}{ccl}
   -1 & \quad & \mbox{for time-like geodesics}
   \\
   0 & \quad & \mbox{for null-geodesics}
   \end{array}\right.\,,
\end{equation}
where $u=ct - z$, $v=ct+z$, $\gamma^u=\gamma^0-\gamma^z$ and $\gamma^v=\gamma^0+\gamma^z$.

Irrespective of the concrete functional dependence of $h^\mathrm{p}$ on the space-time coordinates, all accelerations are zero if $\dot{\gamma}^u=0$. Although $\dot{\gamma}^u=0$ cannot be realized for time-like geodesics, where by definition $-\dot{\gamma^u}\dot{\gamma^v} + h (\dot{\gamma^u})^2 + (\dot{\gamma}^x)^2 + (\dot{\gamma}^y)^2 = -1$, it can be realized for null-geodesics, provided $\dot{\gamma}^x=\dot{\gamma}^y=0$. The corresponding solution reads $\gamma^z(t) = \gamma^z_0 + ct$, $\gamma^x(t)=\gamma^x_0$, $\gamma^y(t)=\gamma^y_0$, which describes propagation on a pulse track parallel, and since for this geodesic the speed $\dot{\gamma}^z$ is always the speed of light, $\dot{\gamma}^z=c$, a massless test particle will never be influenced if co-propagating with the light pulse on a parallel track. For counter-propagating massless test particles, again with $\dot{\gamma}^x=\dot{\gamma}^y=0$, the second solution of the subsidiary condition (\ref{eq:subsed}) resolves into the first order non-linear differential equation $\dot{\gamma}^z=-\frac{1-h^\mathrm{p}}{1+h^\mathrm{p}} c$.  The second solution is, however, only admissible in space time regions where $\partial_x h^\mathrm{p}=\partial_y h^\mathrm{p}=0$, i.e. in regions $I_-$, $I_+$ and $III$. The metric perturbation is zero in regions $I_-$ and $I_+$ and there is no deflection of the massless test particle. In region $III$, the equation of motion can be solved as the curve $\gamma^v(\gamma^u)=\kappa\left[\gamma^u\ln(\gamma^u)-(\gamma^u-L)\ln(\gamma^u-L)\right]+const.$, where the parameterization by $\gamma^u$ is well defined because by definition $\dot{\gamma}^u\neq 0$ for the second null-solution.
In region $III$, a massless test particle counter-propagating with respect to the pulse parallel to the $z$-axis seems to be decelerated and later accelerated again in its propagation direction. 

As this example of a seemingly varying speed of light shows, we have to be very careful with the interpretation of the geodesic equations (\ref{eq:geodesiceq}) and their solutions. "Rods and clocks" used for measurements in the laboratory consist of matter with kinematics which are governed itself by the metric $g=\eta+h$. Hence, the coordinates $(t,x,y,z)$ that were connected to measurements of time and space in the unperturbed spacetime do not represent measurements in the perturbed spacetime. Only after taking the effect of the metric perturbation on "rods and clocks" into account, the geodesic equations (\ref{eq:geodesiceq}) can be interpreted. In the following section, we will show that there is actually no physical effect of the metric perturbation in region $III$.

\section{Curvature and the physical content of the metric perturbation}
\label{sec:curv}

\begin{figure}[h]
\hspace{1cm}
\includegraphics[width=10cm,angle=0]{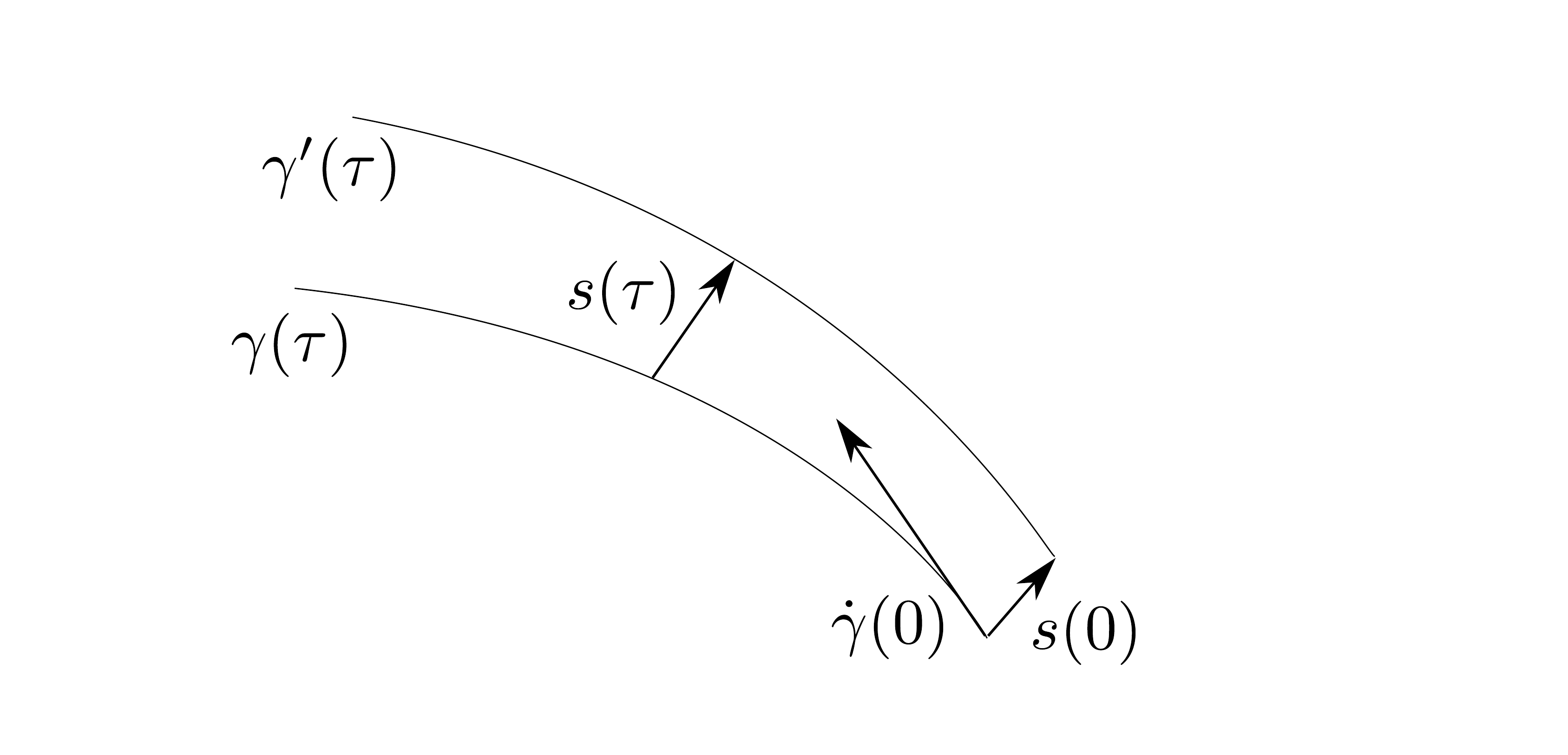}
\caption{The distance between two geodesics changes due to curvature. 
 \label{fig:geodesic_dev}}
\end{figure}
To find out if locally (in a region) any physical effect can arise from a metric perturbation, we can use the Riemann curvature tensor ${R^\mu}_{\nu\rho\sigma}$. If the Riemann tensor vanishes in a region, the spacetime in this region is flat (see section 13.9 in \cite{Straumann2004} and proposition 2.11 of \cite{Atiyah1973}), i.e. a coordinate transformation can be found in which the metric $g$ looks like the Minkowski metric $\mathrm{diag}(-1,1,1,1)$. These coordinates represent measurements with "rods and clocks" in the perturbed spacetime and all test particles move on straight lines in these coordinates. Hence, no physical effects can arise if the Riemann tensor vanishes.

The Riemann tensor ${R^{\mu}}_{\rho\sigma\alpha}$ has a direct physical and geometrical interpretation as it appears naturally in the geodesic deviation equation for the relative acceleration between two infinitesimally close geodesics $\gamma(\lambda)$ and $\gamma'(\lambda)=\gamma(\lambda)+s(\lambda)$ parameterized by $\lambda$ (see figure \ref{fig:geodesic_dev}):
\begin{equation}\label{eq:geodesicdev}
	a^\mu=\frac{D^2s^\mu}{d\lambda^2}={R^{\mu}}_{\rho\sigma\alpha}(x)\dot{\gamma}^\rho \dot{\gamma}^\sigma s^\alpha\,,
\end{equation}
where $s$ is the separation vector between the geodesics and $D/d\lambda=\dot\gamma^\mu \nabla_\mu$ is the covariant derivative along the geodesic $\gamma(\lambda)$. Equation (\ref{eq:geodesicdev}) can be interpreted as the effect of tidal forces on neighboring test particle.

In the following, we will only use the fully covariant version of the curvature tensor $R_{\nu\rho\sigma\alpha}:=\eta_{\nu\mu}{R^\mu}_{\rho\sigma\alpha}$. In first order in the metric perturbation $h_{\mu\nu}$, it takes the form
\begin{equation}\label{eq:defriemann}
	R_{\nu\rho\sigma\alpha}=\frac{1}{2}\left(\partial_{\rho}\partial_{\sigma}h_{\nu\alpha}-\partial_{\nu}\partial_{\sigma}h_{\rho\alpha}-\partial_{\rho}\partial_{\alpha}h_{\nu\sigma}+\partial_{\alpha}\partial_{\nu}h_{\rho\sigma}\right)\,.
\end{equation}
$R_{\nu\rho\sigma\alpha}$ has only $20$ independent components. This can be seen from its symmetries $R_{\nu\rho\sigma\alpha}=-R_{\rho\nu\sigma\alpha}=-R_{\nu\rho\alpha\sigma}$ and $R_{\nu\rho\sigma\alpha}=R_{\sigma\alpha\nu\rho}$ and the Bianchi identity $R_{\nu\rho\sigma\alpha}+R_{\nu\sigma\alpha\rho}+R_{\nu\alpha\rho\sigma}=0$ it fulfills. Due to its symmetries, the Riemann tensor is invariant under linearized coordinate transformations $x^\mu\rightarrow x^\mu+\xi^\mu$ where $\xi^\mu$ is assumed to be small, and it is only considered in first order.

\begin{figure}[h]
\hspace{2.1cm}
\includegraphics[width=6cm,angle=0]{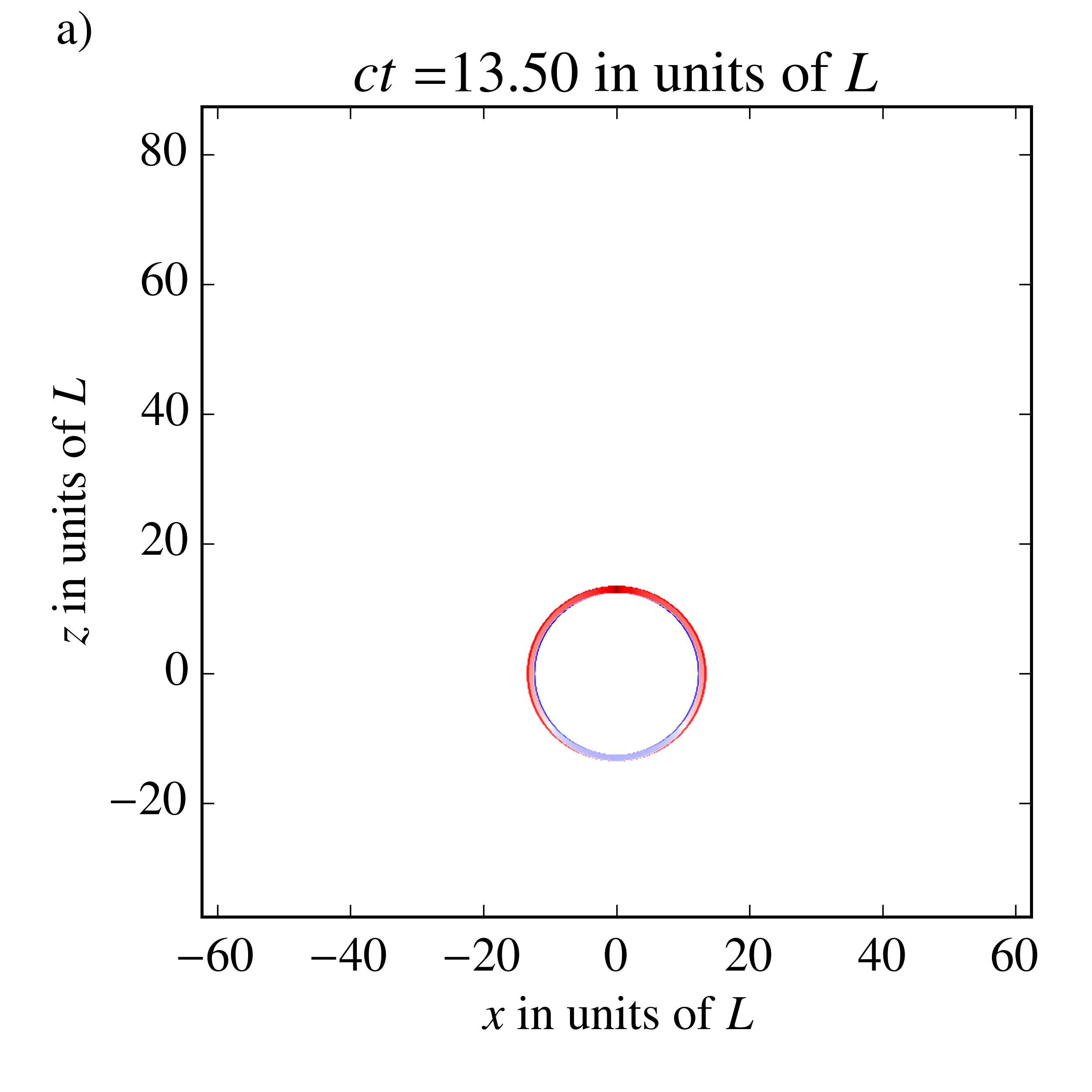}
\includegraphics[width=6cm,angle=0]{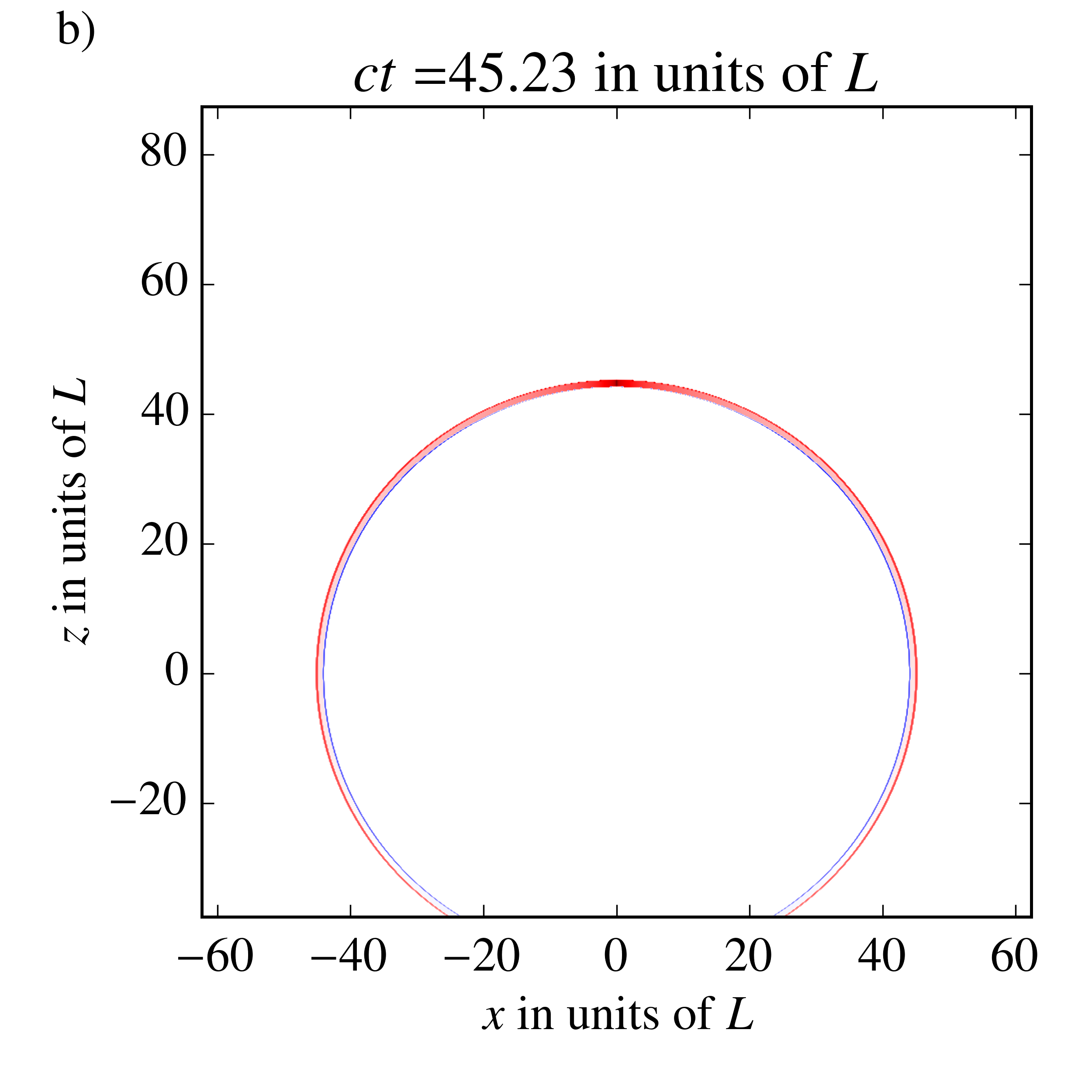}

\hspace{2.1cm}
\includegraphics[width=6cm,angle=0]{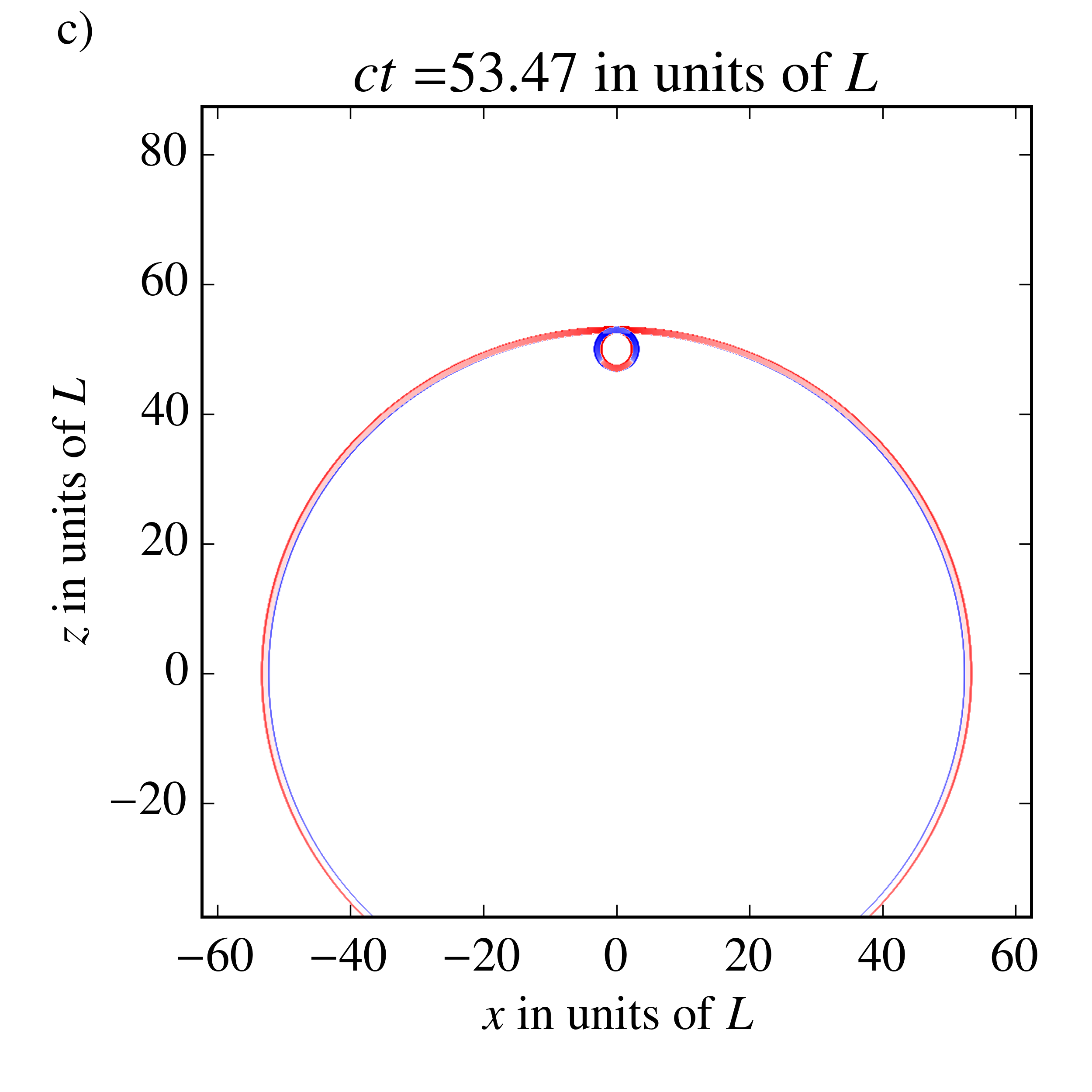}
\includegraphics[width=6cm,angle=0]{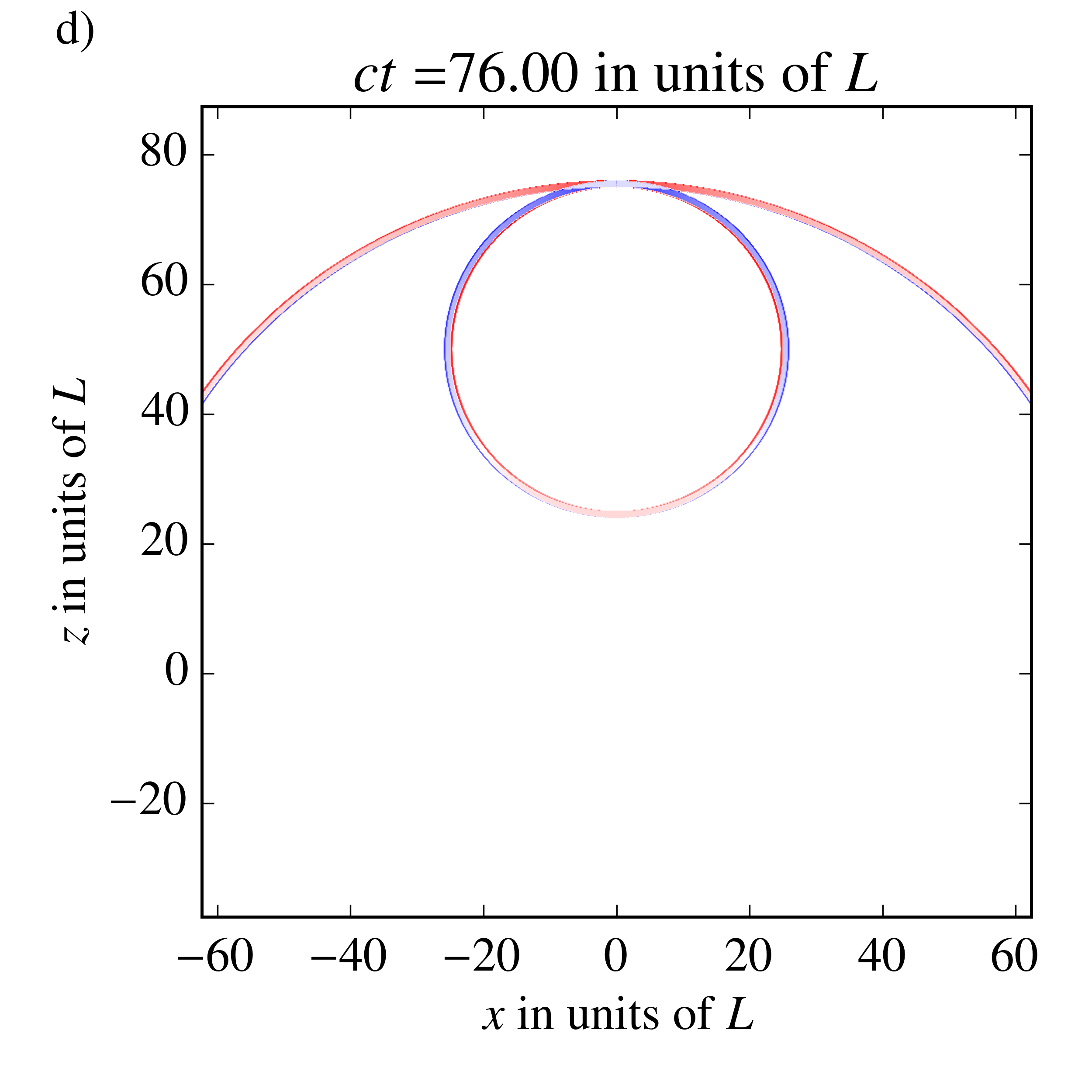}
\caption{\label{fig:gradgradval} The plots show the curvature component $R^\mathrm{p}_{0x0x}$ for the metric perturbation $h^\mathrm{p}_{\mu\nu}$ induced by a laser pulse in the coordinates $(ct,x,y,z)$ in the $(x,z)$-plane for different times $t$. The logarithm of value of $R^\mathrm{p}_{0x0x}$ is encoded in the opacity of the color. Red is a negative value of $R^\mathrm{p}_{0x0x}$ and blue a positive value. White stands for zero.}
\end{figure}
In the following, we will give the components of the curvature tensor due to the light pulse, $R^\mathrm{p}_{\nu\rho\sigma\alpha}$, in terms of derivatives of $h^\mathrm{p}$. Of the $20$ independent components, most turn out to be zero. We find for the only non-zero independent components
\begin{eqnarray}\label{eq:Rallleft1}
	R^\mathrm{p}_{0z0z}&=&-\frac{1}{2}\,(\partial_0+\partial_z)^2 h^\mathrm{p}\\
	R^\mathrm{p}_{0z0i}=-R^\mathrm{p}_{0zzi}&=&-\frac{1}{2} \partial_i(\partial_0+\partial_z)h^\mathrm{p}\\
	R^\mathrm{p}_{0i0j}=R^\mathrm{p}_{zizj}=-R^\mathrm{p}_{0izj}&=&-\frac{1}{2} \partial_i \partial_j h^\mathrm{p}\,,
\end{eqnarray}
where the indices $i$ and $j$ can be $x$ and $y$. See figure \ref{fig:gradgradval} for plots of the curvature component $R^p_{0x0x}$.

In region $III$, $(\partial_0+\partial_z)h^\mathrm{p}=0=\partial_j h^\mathrm{p}$ and $R^\mathrm{p}_{\nu\rho\sigma\alpha}$ vanishes. Hence, the metric perturbation can be removed by a linearized coordinate transformation (see Appendix B for an explicit example of such a transformation). This means, in particular, that the metric perturbation produces a physical effect only in region $II$, $IV$ and $V$ due to the emission, absorption and emission and absorption, respectively. The localization of the gravitational field induced by the laser pulse is due to its motion with the speed of light. If it would move with $v<c$, its effect would be extended over the whole region that is causally connected to the point of its emission as can be seen in \cite{Scully1979} by Scully. The same situation arises for the electromagnetic field induced by a massless charged particle \cite{Azzurli2014}. The localization of the gravitational field of a light pulse can be shown under much more general assumptions than those we use in our one-dimensional model for the pulse, e.g. for a non-diverging, luminal propagating three-dimensional pulse with localized emission and absorption.

\subsection{Curvature very close to the pulse trajectory and far from its emission}

By direct calculation of the curvature components (see Appendix C), we find that, for $\rho/r\rightarrow 0$, the components $R^\mathrm{p}_{0z0z}$, $R^\mathrm{p}_{0z0i}$ and $R^\mathrm{p}_{0zzi}$ decay while the components $R^\mathrm{p}_{0i0j}$, $R^\mathrm{p}_{zizj}$ and $R^\mathrm{p}_{0izj}$ go like $1/\rho^2$. The latter property means also that the influence of the source of the pulse can be neglected in agreement with the argument presented at the end of section \ref{sec:pulse}. We find that in region $II$
\begin{eqnarray}\label{eq:Rremainder}
	R^\mathrm{p}_{0i0j}=R^\mathrm{p}_{zizj}=-R^\mathrm{p}_{0izj}=\frac{4GA}{c^4} \, u(z-ct)\frac{1}{\rho^2}\left(\delta_{ij}-2\frac{x^i x^j}{\rho^2}\right) \,,
\end{eqnarray}
while $R^\mathrm{p}_{0z0z}=R^\mathrm{p}_{0z0i}=-R^\mathrm{p}_{0zzi}=0$.
The curvature in (\ref{eq:Rremainder}) corresponds to the metric perturbation $h_{00}^\mathrm{p} = h_{zz}^\mathrm{p} = -h_{0z}^\mathrm{p}=-h_{z0}^\mathrm{p}=-8GA/c^4\,  u(z-ct) \ln(\frac{\rho}{\alpha})$ with $\alpha$ a constant of the same dimension as $\rho$. This is a special case of the general pp-wave solution of the full Einstein equations for a massless one-dimensional fluid derived in \cite{Bonnor1969}.

With the only non-zero curvature components given in equation (\ref{eq:Rremainder}), the geodesic deviation equation (\ref{eq:geodesicdev}) can be rewritten as a vector equation for the vectors $\vec a:=(a^0+a^z,a^x,a^y)$, $\hat\rho:=(0,x/\rho,y/\rho)$, $\hat\vartheta:=(0,-y/\rho,x/\rho)$, $\vec{\dot\gamma}:=(-\dot\gamma^0+\dot\gamma^z,\dot\gamma^x,\dot\gamma^y)$ and $\vec s:=(-s^0+s^z,s^x,s^y)$:
\begin{equation}\label{eq:tidal}
	\vec a= \frac{4GA}{c^4}\frac{ u(z-ct)}{\rho^2}\left[ ((\hat\rho\times\vec{\dot\gamma})\cdot \vec s)(\hat\rho\times\vec{\dot\gamma})- ((\hat\vartheta\times\vec{\dot\gamma})\cdot \vec s)(\hat\vartheta\times\vec{\dot\gamma})\right]\,,
\end{equation}
while $a^0-a^z=0$.
We see immediately that there are no tidal forces seen by massless co-propagating particles since then $\vec{\dot\gamma}=0$. 

To get a feeling for the magnitude of the tidal forces due to a laser pulse, we can compare the curvature component $R^\mathrm{p}_{0x0x}$ induced by a laser pulse of circular polarization with that of a gravitational quadrupole wave. We can compare them directly because the components of the curvature tensor are invariant under linearized coordinate transformations in first order in the metric perturbation. In the widely used transverse traceless gauge, the curvature component $R^\mathrm{qp}_{0x0x}$ of a gravitational quadrupole wave is given as the second time derivative of the amplitude $h_+$. For a wave of angular frequency $\omega$ we have $R^\mathrm{qp}_{0x0x}= h_+\omega^2/c^2$. The curvature component due to the laser pulse at a point in the $x$-$z$-plane is $R^\mathrm{p}_{0x0x}=\kappa/\rho^2$. For a laser of pulse power $P$, we have $\kappa=4GP/c^5$. Hence, at a distance of $1\mathrm{cm}$ to the pulse trajectory, a laser with $10^{15}$ watt - like the one at the National Ignition Facility - induces curvature of the same order as the curvature of a gravitational wave of angular frequency $\omega=10^3\mathrm{Hz}$ and amplitude $h_+=10^{-22}$. This is of the same order as the strain induced by gravitational waves in this frequency range expected from astronomical sources \cite{Sathyaprakash2009}.

\section{Deflection of test particles}
\label{sec:defltow}

In this section, we want to investigate the effect of the gravitational field of a laser pulse on nearby test particles in more detail keeping in mind the results of section \ref{sec:curv}. Note that due to the rotational symmetry of the metric perturbation around the $z$-axis, there cannot be any non-zero deflection in the azimutal direction (see equation \ref{eq:geodesicdevexpl}), and we can restrict all our considerations to the $x$-$z$-plane. We will only consider the deflection of test particles initially at rest or running parallel to the $z$-axis.

Now, we have to find a coordinate independent measure for the deflection of test particles that can be evaluated in the laboratory. An operational prescription of measuring spatial distances is given by sending light back and forth between separated mirrors. However, this radar distance can be only defined for distances that can be traversed by light before the gravitational field changes significantly. For region $II$, $IV$ and $V$, this implies a maximal length scale for distance measurements at the order of $L$.
Luckily, all test particles except the co-propagating, massless one, traverse regions $II$, $IV$ and $V$ in finite time, and we can evaluate their accumulated deflection in the flat regions $III$ and $I_+$, respectively. 
For infinitesimally separated points, this radar distance coincides with the proper distance which is defined as $\sqrt{g_{\mu\nu}s^\mu s^\nu}$ for the separation vector $s$ between the points. 

Then we have to find a connection between the deflection of test particles measured in the coordinates $(ct,x,y,z)$ and the deflection  measured in proper distance. The latter turns out to be mostly towards or away from the $z$-axis for $\rho/r\ll 1$ which we identified as a necessary condition for neglecting the contributions of emitter and absorber in section \ref{sec:pulse}.
For $\rho/r\ll 1$, the curvature components $R^\mathrm{p}_{0i0j}$, $R^\mathrm{p}_{zizj}$ and $R^\mathrm{p}_{0izj}$ become dominant as we have shown in the course of deriving equation (\ref{eq:Rremainder}) (see also Appendix C). This dominance means that the only significant relative acceleration between two neighboring geodesics with $\dot\gamma^x(\lambda_0)=0$ for some value of the curve parameter $\lambda_0$ is $a^x(\lambda_0)$ as can be verified directly from the geodesic deviation equation (\ref{eq:geodesicdev}). 

The relative acceleration provides information about the proper distance between two neighboring geodesics. For parallel geodesics, the second derivative of the proper distance for the curve parameter $\lambda$ is the projection of the relative acceleration along $s$, i.e. $d^2/d\lambda^2 \,\sqrt{g_{\mu\nu} s^\mu s^\nu} = g_{\mu\nu} a^\mu \hat{s}^\nu$ where $\hat{s}^\nu = s^\nu/\sqrt{g_{\mu\nu} s^\mu s^\nu}$ is the normalized distance vector. In the case of neighboring geodesics with $\dot\gamma^x(\lambda_0)=0$ for some value of the curve parameter $\lambda_0$, $\gamma^x(\lambda_0)>0$ and distance vector $s(\lambda_0)$ in the $x$-direction, the acceleration of change in the proper distance becomes $\ddot s^x(\lambda_0) = d^2/d\lambda^2 \sqrt{(\eta+h^\mathrm{p})_{\mu\nu} s^\mu s^\nu}(\lambda_0) =  a^x (\lambda_0) $ since $h^\mathrm{p}_{x\nu}=0$ for all $\nu$. 

Hence, the absolute acceleration towards or away from the $z$-axis is well defined, and we obtain it by integrating $\ddot s^x$ along the $x$-axis. It is $\ddot\gamma^x$ for the trajectory of a test particle governed by the geodesic equation (\ref{eq:geodesiceq})
\begin{equation}\label{eq:x2dev}
	\ddot{\gamma}^x=\frac{1}{2}(\dot \gamma^0 -\dot \gamma^z)^2\partial_x h^\mathrm{p}\,.
\end{equation}
$\ddot{\gamma}^x$ and the change of time measurement due to the metric perturbation are both of first order in the metric perturbation $h^\mathrm{p}$. Hence, in first order in $h^\mathrm{p}$, $\ddot{\gamma}^x$ and $\dot{\gamma}^x$ can be interpreted as the acceleration and the velocity of the test particle in the $x$ direction, respectively.

\begin{figure}[h]
\hspace{2.1cm}
\includegraphics[width=6cm,angle=0]{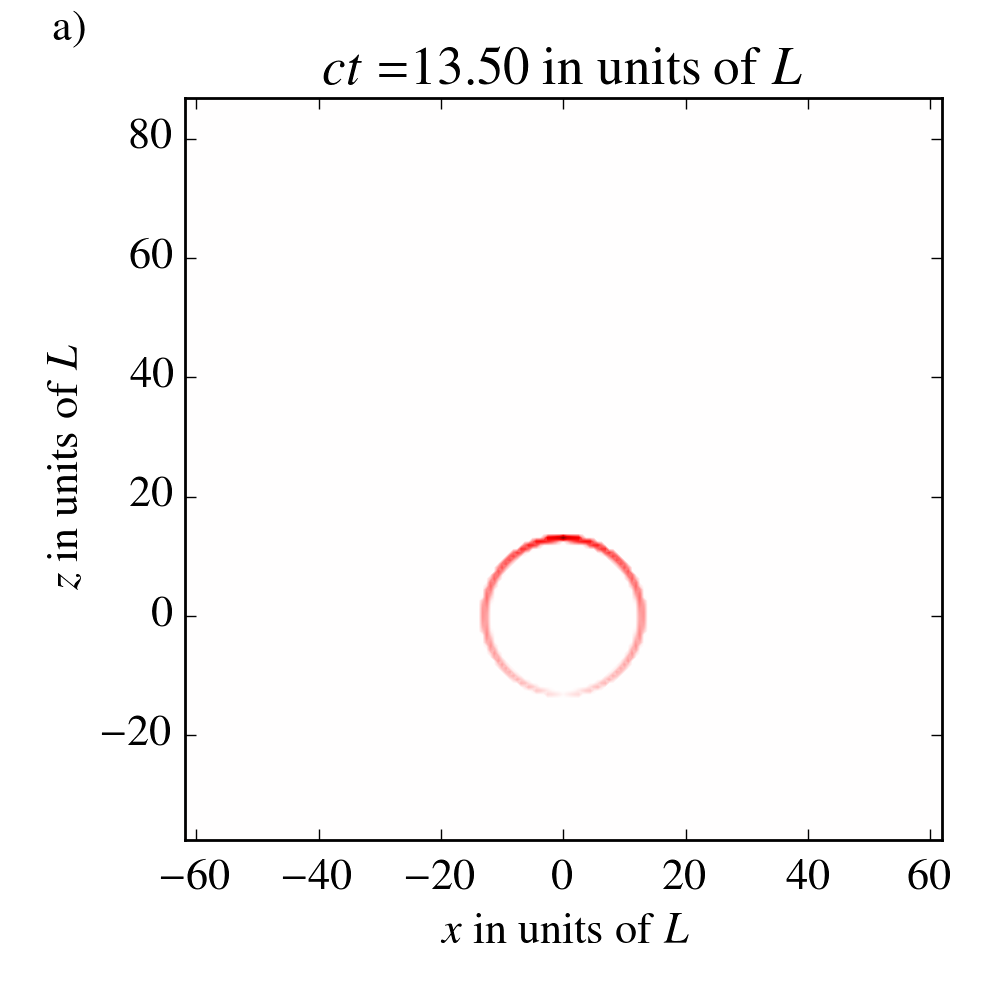}
\includegraphics[width=6cm,angle=0]{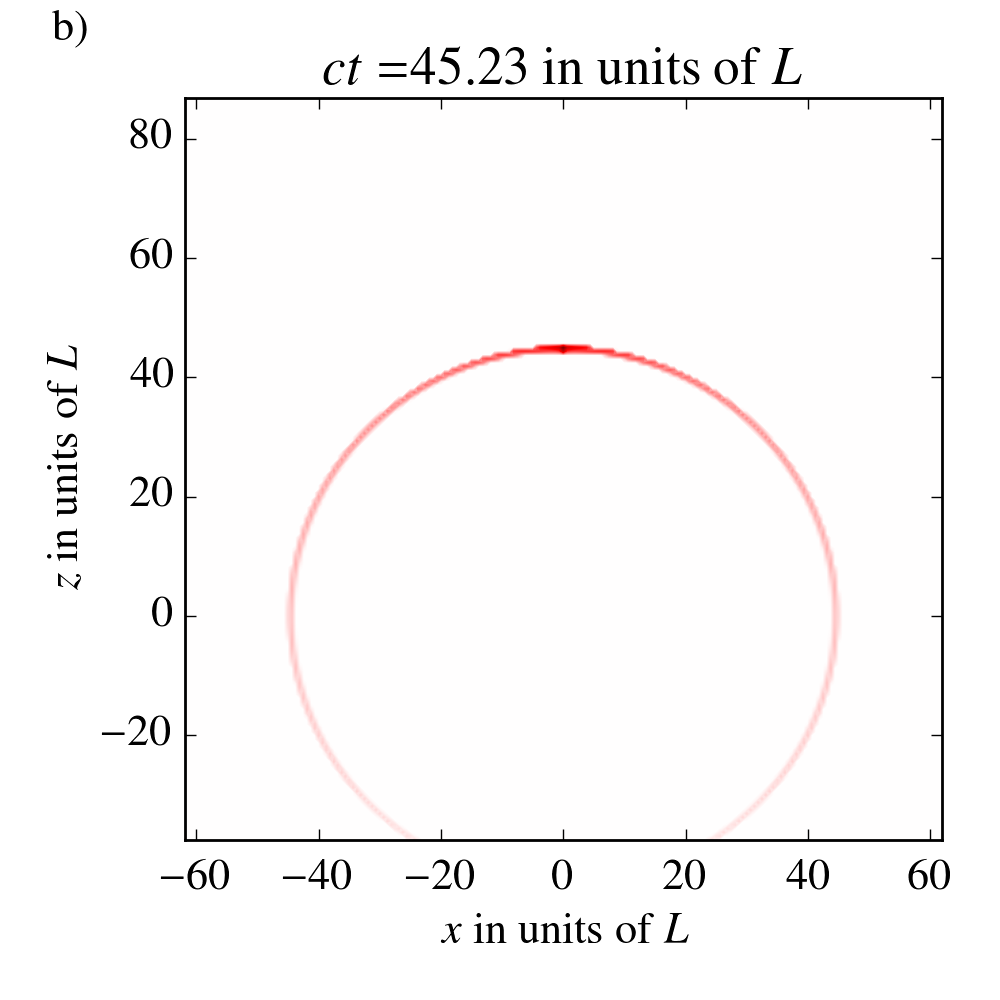}

\hspace{2.1cm}
\includegraphics[width=6cm,angle=0]{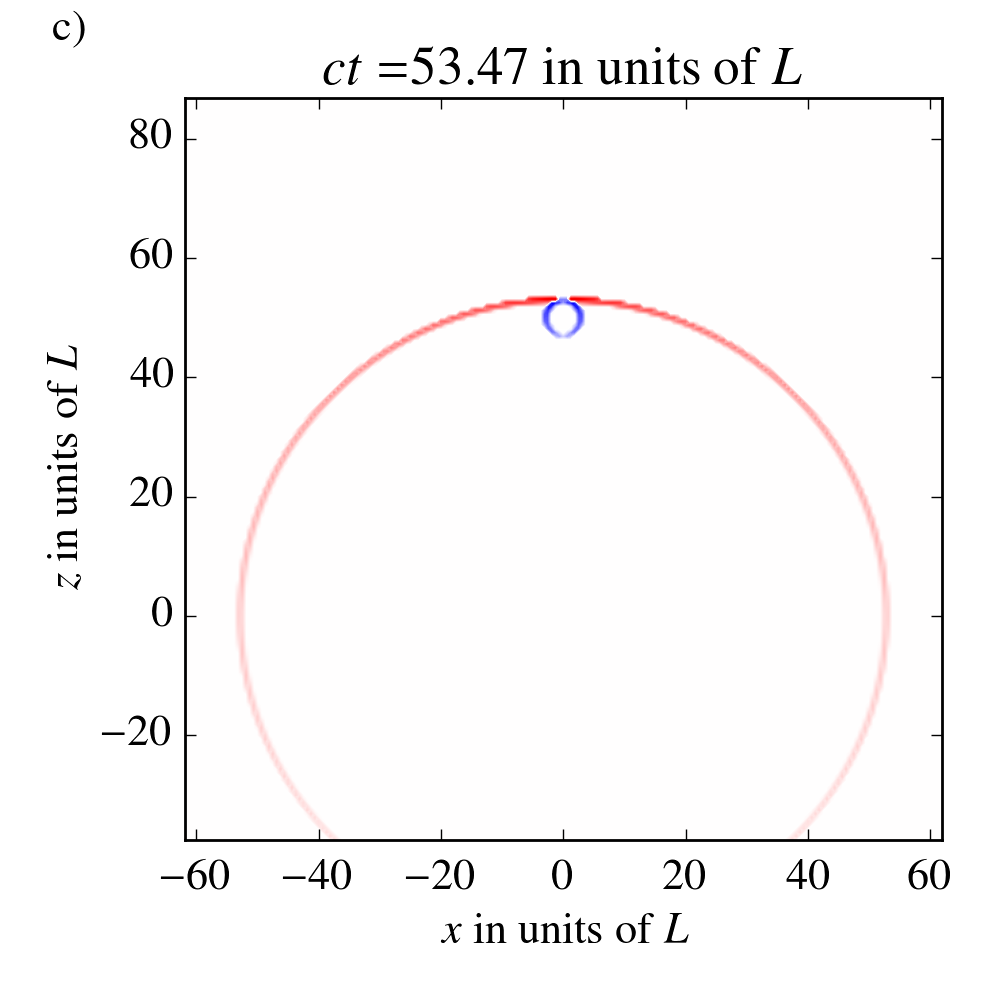}
\includegraphics[width=6cm,angle=0]{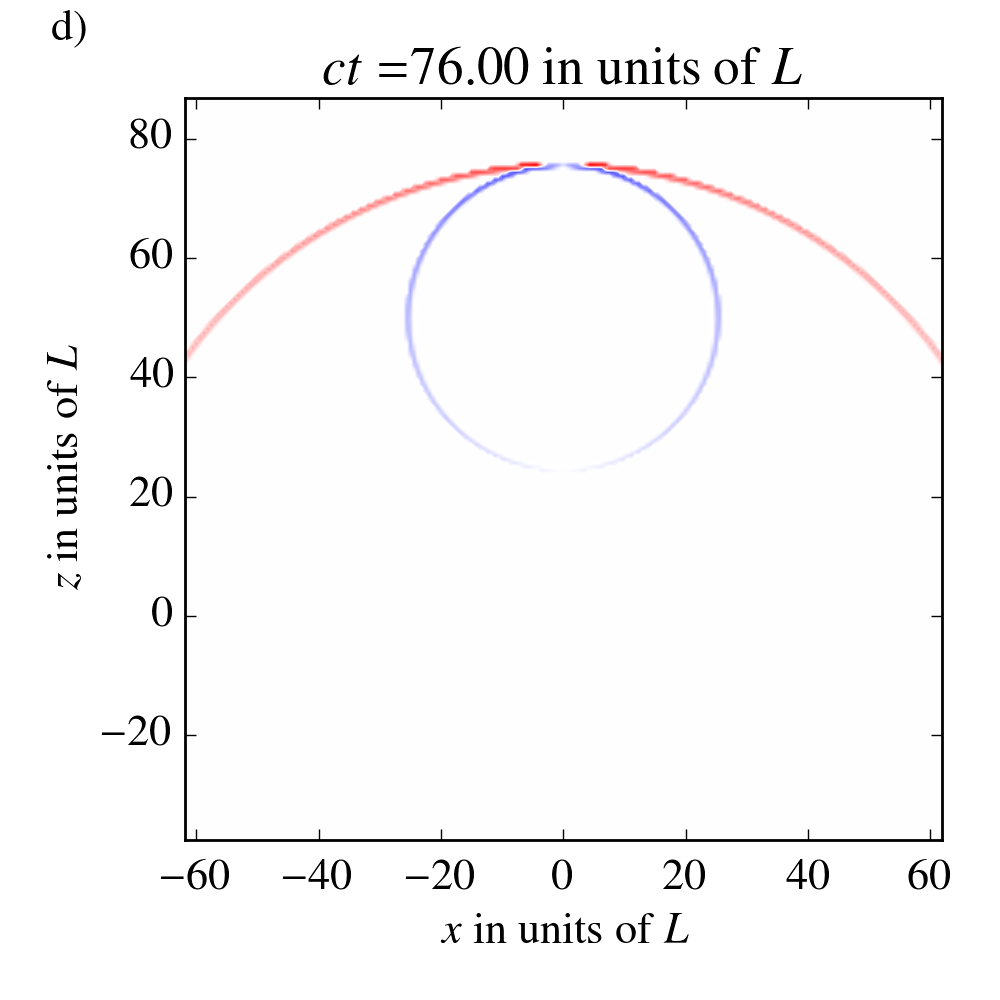}
\caption{\label{fig:gradhval}The plots show the acceleration $\ddot \gamma^x$  towards/away from the $z$-axis experienced by a test particle (with $\gamma^x=\gamma^y=0$) due the laser pulse in the $(x,z)$-plane for different times $t$. The logarithm of $|\partial_x h^\mathrm{p}|$ which is proportional to the force (see equation (\ref{eq:x2dev})) is encoded in the opacity of the color. Red corresponds to an attraction and blue to a repulsion. White stands for a vanishing acceleration.}
\end{figure}
The factor $\partial_x h^\mathrm{p}$ follows from equation (\ref{eq:hint}) and equation (\ref{eq:regions}) as
\begin{eqnarray}\label{eq:partial2h}
 \partial_x h^\mathrm{p}(t,x,y,z)&=& \frac{4GA}{c^4}\left(\chi_{L}(ct-r) u\left(r - ct\right)\partial_x\ln(r-z)-\right.\\
	 &&\left. - \chi_{L}(ct-r_D) u\left(r_D - ct\right)\partial_x\ln(r_D - z)\right) \nonumber
\end{eqnarray}
where $\vec r=(x,y,z)$, $r=|\vec r|$, $r_D=|\vec r - (0,0,D)| + D$ and  $\chi_{L}$ is the characteristic function which is zero outside the interval $[0,L]$ and $D$ is the distance between emitter and absorber.

Let us have a closer look at the two terms in equation (\ref{eq:partial2h}). The first term is zero outside of an expanding spherical shell of width $L$ centered at the origin. It only depends on the source at the retarded time $ct-r$ at the origin. So, it corresponds to the emission process. It gives an attraction towards the $z$-axis.
The second term is zero outside of an expanding spherical shell of width $L$ centered at the point $(x,y,z)=(0,0,D)$. It only depends on the source at the retarded time $ct-r_D$ at the point $(x',y',z')=(0,0,D)$. Hence, it corresponds to the absorption process. It gives a repulsion from the $z$-axis.

\subsection{A test particle at rest}

Let us have a look at the particular case of a test particle at rest in the laboratory frame, i.e. $\dot \gamma^0=c$ and $\dot \gamma^z=0$. The equation (\ref{eq:x2dev}) tells us that the acceleration is
\begin{equation}\label{eq:force}
	\ddot \gamma^x =\frac{c^2}{2}\partial_x h^\mathrm{p}\,.
\end{equation}
For example, on the $z$-axis, i.e. at $x=y=0$, this test particle would first experience an acceleration towards the $z$-axis at $t=\frac{1}{c}z$ for a time span $\frac{1}{c}L$. At $t=\frac{1}{c}(|z-D|+D)$, the test particle would experience an acceleration away from the $z$-axis for a time span of $\frac{1}{c}L$.

\begin{figure}[h]
\hspace{2.1cm}
\includegraphics[width=9cm,angle=0]{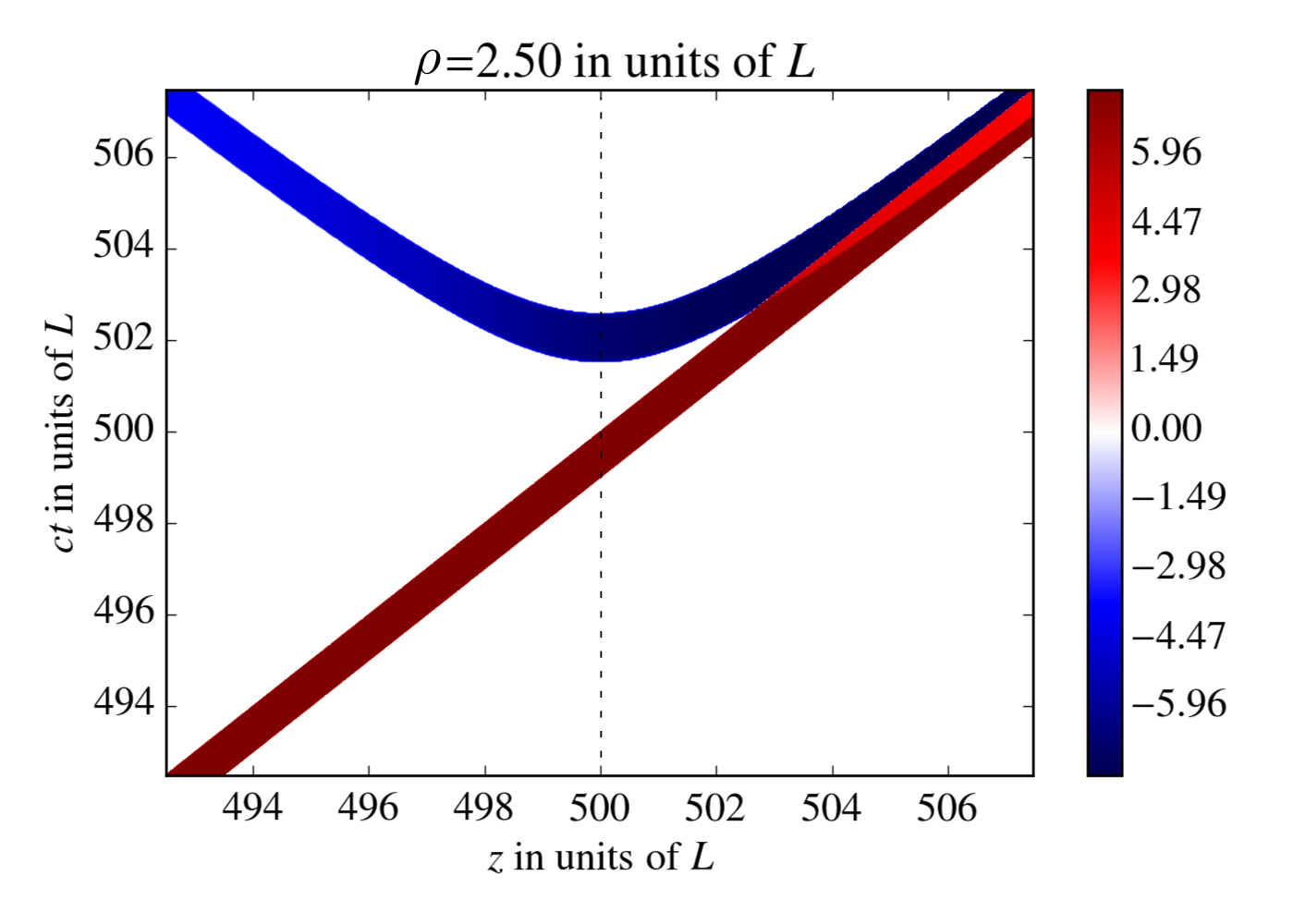}

\hspace{2.1cm}
\includegraphics[width=9cm,angle=0]{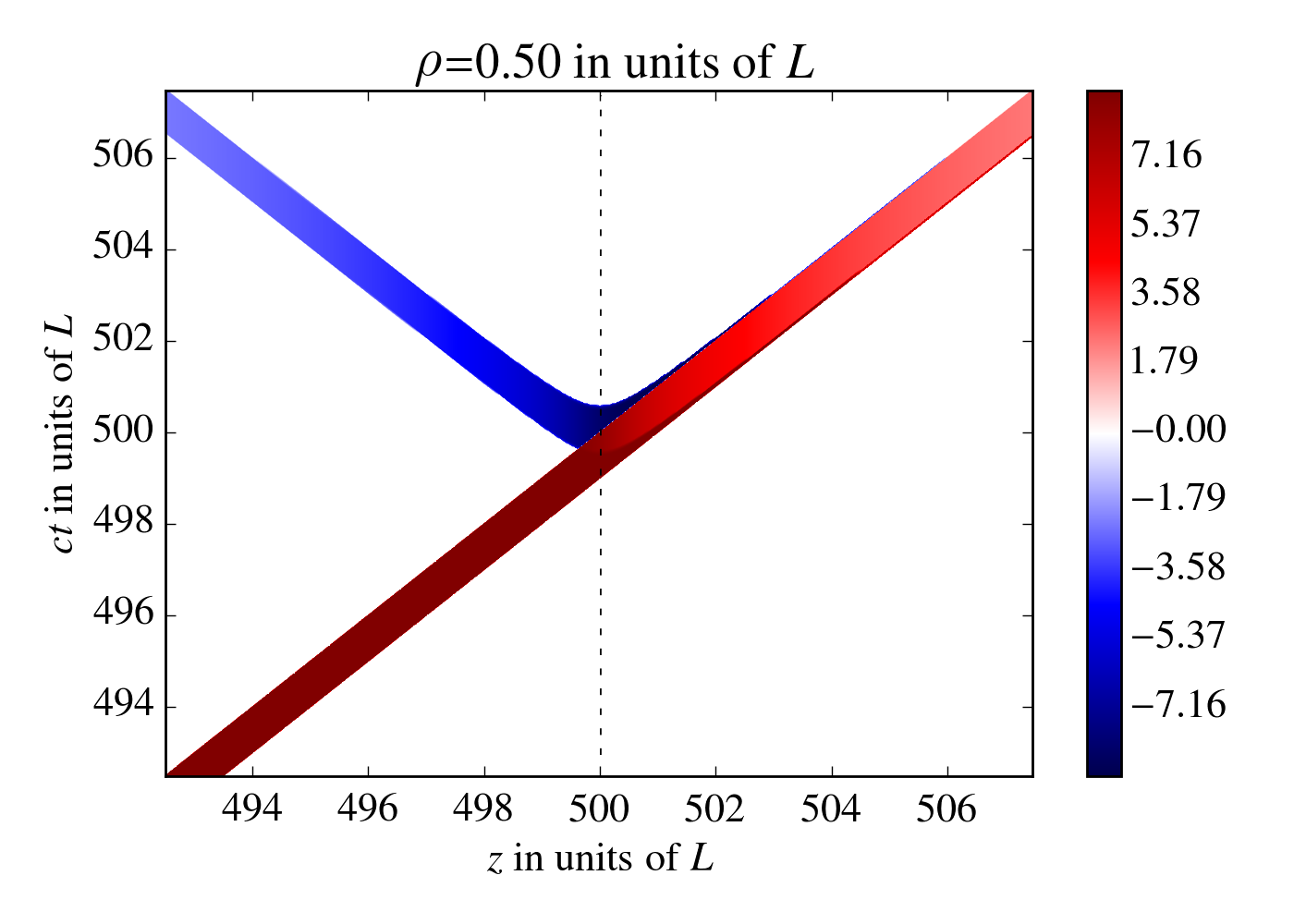}
\caption{\label{fig:grad_ct_x} Acceleration (red: attraction; blue: repulsion; logarithmic scale) experienced by a test particle at rest at different $z$-positions for different times $t$ after the emission of a circular polarized laser pulse of length $L$. The point of absorption is at $z=D=500L$. The acceleration is given in units of $c^2\frac{\kappa}{2}$. Upper: distance $\rho$ of the test particle to the $z$-axis is $2.5L$; Lower: $\rho=0.5L$.}
\end{figure}

In figure (\ref{fig:grad_ct_x}), there are plots of the acceleration in the $x$-direction experienced by test particle at rest at different positions parallel to the $z$-axis at two different distances $\rho$ from the $z$-axis for different times after the emission of a circularly polarized laser pulse.
A test particle at $z < D$ will always first experience an attractive force and later a much weaker repulsive force. A test particle with $z > D$ but a distance $\rho$ from the $z$ axis at least of the order of the length of the laser pulse will experience first an attractive force and later a weaker repulsive force. In the latter case, the delay of the two forces and their difference in magnitude decrease for increasing $z$. The total deflection in the $x$-direction vanishes for $z\rightarrow \infty$. 

In region $II$ (the causal past of the emission process), for points in the $x$-$z$-plane, for circular polarization, $\rho/r\ll 1$ and positive $z$, we get from equation (\ref{eq:hintcirc}) the result $\partial_x h^\mathrm{p}\approx - 2\kappa/x$ which leads to
\begin{equation}\label{eq:forcecirc}
	\ddot \gamma^\rho\approx -c^2\frac{\kappa}{x}\,.
\end{equation}
The constant $\kappa=2G\varepsilon_0 E_0^2 Ac^{-4}$ can be rewritten for a laser pulse of power $P$ as
\begin{equation}
	\kappa=\frac{4GP}{c^5}
\end{equation}
The strongest laser pulses available today have a pulse power in the range of $10^{15}\mathrm{W}$. At a distance of $2.5\mathrm{mm}$, this gives the acceleration 
\begin{equation}\label{eq:accNIF}
	\ddot \gamma^x \approx -\frac{4GP}{c^3x}\approx - 10^{-18}\frac{\mathrm{m}}{\mathrm{s}^2}\,.
\end{equation}
We can compare this acceleration to the acceleration experienced by a test particle in the Newtonian potential induced by a small spherical, massive object. At a distance of $r=2.5\mathrm{mm}$, a mass of only $M=10^{-13}\mathrm{g}$ would be necessary to provide the same acceleration as in equation (\ref{eq:accNIF})
\begin{equation}
	\ddot \gamma^r =-\frac{GM}{r^2}\approx - 10^{-18}\frac{\mathrm{m}}{\mathrm{s}^2}\,.
\end{equation}
We see again: the gravitational field of light is too small to be detected in the near future.

\subsection{A massless test particle}

Now, let us have a look at two different cases for the deflection of massless test particles, e.g. photons. First, we assume that the massless test particle moves in the same direction as the laser pulse, the positive $z$-direction. This means $\dot x^0=c=\dot x^z$ and from equation (\ref{eq:x2dev}) follows that the deflection is zero as we found in section \ref{sec:geodesics}.
The second case is that of a massless test particle moving in the negative $z$-direction, i.e. it is counter-propagating. We have $\dot x^0=c=-\dot x^z$, and we find from equation (\ref{eq:x2dev})
\begin{equation}\label{eq:deflection}
	\ddot{\gamma}^x=2c^2\partial_x h^\mathrm{p}\,.
\end{equation}
equation (\ref{eq:deflection}) results in a non-zero acceleration in the $x$-direction which corresponds to a nonzero gravitational force acting on the massless test particle, and we recover the result of \cite{Tolman1931}. The acceleration (\ref{eq:deflection}) is four times that experienced by a test particle at rest (compare with equation (\ref{eq:force})). Since the massless test particle moves with $c$, however, it will need a different time span to pass through the regions $II$, $IV$ and $V$ than the particle at rest. For $\rho/r\ll 1$ it will only experience the acceleration (\ref{eq:deflection}) for $L/2c$ while the particle at rest will be accelerated for $L/c$. Hence, the total deflection of a counter-propagating, massless test particle will be only by about a factor $2$ larger than that of a particle at rest.

\subsection{2D Newtonian gravity}

It is interesting to notice that the gravitational effect on test particles in region $II$ resembles Newtonian gravity in two spatial dimensions. This similarity is easily seen for $\rho/r\rightarrow 0$ when the curvature takes the form (\ref{eq:Rremainder}) which corresponds to the metric perturbation $h^\mathrm{p}=h_{00}^\mathrm{p} = h_{zz}^\mathrm{p} = -h_{0z}^\mathrm{p}=-h_{z0}^\mathrm{p}=-8GA/c^4\,  u(z-ct) \ln(\frac{\rho}{\alpha})$ with $\alpha$ a constant of the same dimension as $\rho$. In particular, $\ln\rho$ is the solution of the Poisson equation in two dimensions with a point source at $\rho=0$ which corresponds to the generalization of the Newtonian potential of a point particle at rest to two dimensions \footnote{The case of finite lifetime corresponds to Newtonian gravity on a two dimensional sphere with a point source and an additional negative surface mass.}.
The derivative of $h^\mathrm{p}$ for $x$ gives  
\begin{equation}\label{eq:x2dev2D}
	\ddot{\gamma}^x=-4GA/c^4(\dot \gamma^0 -\dot \gamma^z)^2 u(z-ct) \frac{1}{x}\,,
\end{equation}
in the $x$-$z$-plane. The acceleration (\ref{eq:x2dev2D}) points towards the pulse and is proportional the inverse distance to the pulse like we would expect from 2D Newtonian gravity. 
Like in Newtonian gravity, the gravitational effect would be interpreted as instantaneous if the pulse and not the emission would be seen as the source of the gravitational field since all points of a given $x$-$y$-plane seem to be effected simultaneously by the pulse when the $x$-$y$-plane passes region $II$.

\section{Conclusions}
\label{sec:conclusions}

We introduced a model for a laser pulse of finite lifetime following the article \cite{Tolman1931} by Tolman et al., and derived the corresponding gravitational field as a retarded potential in the framework of linearized gravity. We argued that the gravitational fields of emitter and absorber can be neglected for the distance $\rho$ to the pulse small in comparison to the distance $r$ to emitter and absorber. 

We found that the gravitational field shows oscillations of half the wavelength of the electromagnetic field corresponding to the pulse if it is linearly polarized. These oscillations are not present for circular polarization. In both cases, the gravitational field is independent of the orientation of the polarization.

We analysed the gravitational effect of the pulse on nearby test particles, and we found that it is only due to the pulse emission and absorption as concluded in \cite{Bonnor2009}. In \cite{Scully1979}, the metric perturbation for a laser pulse moving with speed $v<c$ is  derived. In contrast to our case, test particles witness gravitational effects in the whole region causally connected with the timeline of the laser pulse propagating with $v<c$. Hence, the localization of the gravitational effect is due to the luminal motion of the laser pulse. A similar situation arises in electrodynamics with massless charges \cite{Azzurli2014}.

The width of a realistic laser pulse will change during its propagation if it is not propagating in a waveguide. However, if we assume the overall change of width to be small, the effect of the spreading should be much smaller than the effect of the emission and absorption. This question may be investigated somewhere else.

We showed that for $\rho/r\ll 1$ the deflection of test particles moving parallel to the $z$-axis is mostly transversal to the trajectory of the pulse. The emission of the pulse induces an attractive effect while the absorption induces a repulsive effect. Hence, a test particle at rest will first be deflected towards the trajectory of the pulse and later repelled. We recovered the result of \cite{Tolman1931} that a massless test particle is not effected by the pulse if it is co-propagating with the pulse while a counter-propagating massless test particle experiences an acceleration four times stronger than that experienced by a particle at rest.

For $\rho/r\rightarrow 0$, we found that the gravitational field converges to a plane fronted parallel propagating gravitational wave (pp-wave) recovering the metric presented in \cite{Bonnor1969}. We calculated the resulting strain of spacetime for a hypothetical peta-watt laser at a distance of $1\mathrm{mm}$ to the beam line, and we found that it is similar to that induced by gravitational waves with frequency $1\mathrm{kHz}$ from astronomical sources.

In the limit $\rho/r\rightarrow 0$, we worked out the similarity of the gravitational field of a laser pulse with Newtonian gravity in two dimensions. The general case will be discussed somewhere else.

\section{Outlook}

The experimental detection of the gravitational effect of a laser pulse is way out of reach as was argued in section \ref{sec:geodesics} and in \cite{Scully1979} and \cite{Brodin2006}. It was pointed out in \cite{Scully1979}, however, that measuring phase shifts is "thinkable". It would be worthwhile to evaluate phase shifts of a test laser induced by a strong laser field for the case of a pulse traveling with the speed of light. The different gravitational effect on a co-propagating and a counter-propagating test beam/pulse could be used in an interferometric detection scheme. A way to deal with the long traveling distances necessary for tests using deflection or phaseshifts is to change the path. In particular ring lasers could be an alternative \cite{Ji2006}. In such geometries also frame dragging arises \cite{Mallett2000}.

Another possibility is the detection of the spacetime strain (the tidal forces) induced by the laser. Interferometric gravitational wave detectors are to large to be used for these kind of applications because the detector has to be very close to the source. However, gravitational wave detectors on the micrometer scale like the one proposed in \cite{Sabin2014} are small enough and, potentially, close to the necessary range of sensitivity.

It would be also interesting to look closer at the background dependence of the gravitational effect of light: we assumed the background to be Minkowski space. For larger traveling distances, earthbound experiments are performed in a Schwarzschild background, however. One first reference in this direction is \cite{Dray1985} in which expressions for the gravitational field of a point like light pulse were derived in a Schwarzschild background. 

But, the background gravitational field can also significantly alter the behavior of laser pulses via the vacuum polarization. In simple cases the speed of light stays the speed of light due to renormalization but already in the Schwarzschild background the tidal forces lead to vacuum birefringence which means that the speed of light depends on the polarization \cite{Drummond1980}. The kinematics of test particles under these conditions where derived in \cite{Raetzel2011}. However, it stays unclear what that means for the backreaction of particles on the gravitational field. A possible way to the corresponding Einstein equations was presented in \cite{Giesel2012}, however, only for very special cases solutions where derived \cite{Schuller2014}.

We found that the gravitational field of a laser pulse does not depend on its helicity. This is because the gravitational field and the spin of matter - its intrinsic angular momentum - are not coupled in general relativity \cite{Ni2010,Hehl1976} which is necessary to ensure the universality of free fall (the weak equivalence principle) \cite{Mashhoon1999}. There is, however, no obvious reason to distinguish between intrinsic angular momentum and extrinsic angular momentum of matter. Even more, spin angular momentum can be transformed to orbital angular momentum \cite{Bliokh2011,Karimi2014} which couples to the gravitational field in general relativity (it generates frame dragging effects \cite{Strohaber2013}). We can also consider an atom absorbing a photon with circular polarization. During that process, the atom will change its total angular momentum by a spin-flip or a change in the orbital angular momentum of one of its electrons. Do we now treat the orbital angular momentum of the electron as intrinsic or extrinsic? This distinction makes even less sense when considering larger quantum systems. In contrast, the gravitational interaction of particles depends on their spin or helicity in perturbative quantum gravity \cite{Boccaletti1969,Boccaletti1967} which can be seen as an effective field theory for a theory of quantum gravity in the low energy limit \cite{Donoghue2012}. Here, we have obviously another mismatch of predictions on the intersection of quantum physics and general relativity. 

It is possible to couple gravity to spin by removing the condition of zero torsion that is imposed in general relativity. There are several approaches to dynamical spacetime theories with torsion among them Einstein-Cartan-Theory and the Poincaré-Gauge-Theory of gravity \cite{Hehl1976}. However, the electromagnetic field cannot be coupled minimally (via the covariant derivative) to the gravitational field without loosing its gauge invariance \cite{Itin2003}. Hence, the only way to obtain a dependence of the gravitational field on the spin of the electromagnetic field in these theories is the introduction of non-minimal coupling. The resulting modification of the constitutive tensor gives then rise to effects like vacuum birefringence \cite{Itin2003}. It should be interesting to study the implications of these modifications in the vicinity of strong laser pulses.

In \cite{Schucker1990}, the differential cross section for the scattering of two light pulses is derived, and it is shown that, for small angles, the result coincides with the differential cross section for photon-photon scattering in the framework of perturbative quantum gravity \cite{Barker1967,Boccaletti1969}. This result becomes meaningful in the framework of semi-classical gravity where the wave packet of a single photon can be associated with a pulse of electromagnetic radiation. Then the gravitational field we derived could be interpreted as the gravitational field of a single-photon pulse.

It would interesting if the localization of the gravitational effect of light still holds in some framework of quantum gravity.

\section*{Acknowledgments}

DR thanks Ivette Fuentes, Friedrich W. Hehl, Carsten Henkel, Axel Heuer, Maximilian Lock, Robert Marx, Jonas Schm\"ole and Christof Zink for helpful remarks and discussions.

\section*{Appendix A}

For linear polarization, we obtain 
\begin{eqnarray}
\hspace*{-2cm}
	h^\mathrm{p}&=&
	2\kappa\int_{\zeta(a)}^{\zeta(b)} d\zeta\frac{\sin^2\left(\frac{\omega}{c}\left(ct-z-\zeta\right)+\varphi\right)}{\zeta}\\
	\nonumber &=&\kappa\int_{\zeta(a)}^{\zeta(b)} d\zeta\frac{1-\cos \left(2\frac{\omega}{c}(ct-z)+\varphi\right) \cos 2\frac{\omega}{c} \zeta - \sin \left(2\frac{\omega}{c}(ct-z)+\varphi\right) \sin 2\frac{\omega}{c} \zeta}{\zeta}
\end{eqnarray}
We find
\begin{eqnarray}
\hspace*{-2cm}
	h^\mathrm{p}&=&	\kappa \left[\ln(2\frac{\omega}{c}\zeta)-\cos \left(2\frac{\omega}{c}(ct-x)+\varphi\right)\text{Ci}(2\frac{\omega}{c}\zeta)  - \sin \left(2\frac{\omega}{c}(ct-x)+\varphi\right)\text{Si}(2\frac{\omega}{c}\zeta)\right]_{\zeta(a)}^{\zeta(b)} \,
\end{eqnarray}
where $\text{Ci}$ and $\text{Si}$ are the integral-cosine and the integral-sine, respectively (see \cite{Abramowitz1964} p. 231).

\section*{Appendix B}

The metric perturbation in region $III$ can even be gauged away by a linearized coordinate transformation $\tilde x^\mu=x^\mu+\xi^\mu$. For circular polarization, one option is
\begin{eqnarray}\label{eq:xicomplete}
\hspace*{-2cm}
	\nonumber\xi^\mu = \frac{2GA}{c^4}\eta^{\mu\nu}p_\nu \Big[&\Theta(ct-r)\left((ct-z)\ln (ct-z)-(ct-z)\right)-&\\
	\nonumber &-\Theta(ct-L-r)\left((ct-L-z)\ln (ct-L-z)-(ct-L-z)\right)-&\\
	&-(\Theta(ct-r)-\Theta(ct-L-r))((r-z)\ln(r-z)-(r-z)) &\Big]\,.
\end{eqnarray}
with $p_{\mu}=(1,0,0,-1)$. What remains is a metric perturbation $\tilde h_{\mu\nu}=h_{\mu\nu}-\partial_\mu\xi_\nu-\partial_\nu\xi_\mu$ that is zero in region $III$. The coordinate transformation (\ref{eq:xicomplete}) induces a non-zero metric perturbation in region $V$ that must be compensated by a similar linearized coordinate transformation. The appropriate coordinate transformation can be obtained from (\ref{eq:xicomplete}) by shifting $ct$ and $z$ by $D$ and reversing the overall sign.

\section*{Appendix C}

The components of the fully covariant Riemann curvature tensor in region $II$ are
\begin{eqnarray}\label{eq:RremainderD}
\hspace*{-2.5cm}
	\nonumber R^\mathrm{p}_{0z0z}&=&-\frac{2GA}{c^4}(\partial_0+\partial_z)\left(\tilde u(r-ct)\frac{1}{r}\right) \\
\hspace*{-2.5cm}
	&=&\frac{2GA}{c^4}\frac{1}{r^2}\Big[\tilde u(r-ct)\frac{z}{r}+\tilde u'(r-ct)(r-z) \Big]\\
\hspace*{-2.5cm}	\nonumber R^\mathrm{p}_{0z0i}=-R^\mathrm{p}_{0zzi}&=&-\frac{\bar h_0}{2}\partial_i\left(\tilde  u(r-ct)\frac{1}{r} \right)\\
\hspace*{-2.5cm}
	&=&\frac{2GA}{c^4}\frac{1}{r^2}x^i\Big[\tilde u(r-ct)\frac{1}{r} - \tilde u'(r-ct)\Big]\\
\hspace*{-2.5cm}	\nonumber R^\mathrm{p}_{0i0j}=R^\mathrm{p}_{zizj}=-R^\mathrm{p}_{0izj}&=&\frac{2GA}{c^4} \partial_i\left( \tilde u(r-ct)\frac{x^j}{r(r-z)}\right) \\
\hspace*{-2.5cm}
	 &=&\frac{2GA}{c^4}\frac{1}{r(r-z)}\left(\tilde u(r-ct)\left(\delta_{ij}-\frac{x^i x^j}{r^2}\frac{2r-z}{r-z}\right)+\tilde u'(r-ct)\frac{x^ix^j}{r}\right) \,,
\end{eqnarray}
where $\tilde u(r-ct)=\chi_L(r-ct) u(r-ct)$. 

In the following, we call the first three independent components in equation (\ref{eq:Rallleft1}) longitudinal curvature components referring to the majority of their indices being $0$ or $z$ while the tangent vector to the trajectory of our pulse has non-zero components $0$ and $z$. In contrast, we will call the last three components in equation (\ref{eq:Rallleft1}) transversal. 

We compare the terms containing $u$ and those containing $u'$ separately. We find that, because of the different factors containing $(r-z)$, $r$, $x^i$ and $z$ in various ways, the transversal curvature components are much larger than the longitudinal components if $\rho\ll r$, i.e. if the distance to the trajectory of the pulse is much smaller than the distance to the point of emission.
In particular, the former become independent of $r$ for $\rho \ll r$ and we can safely assume that the transversal components dominate the corresponding components $R_{0i0j}$, $R_{zizj}$ and $R_{0izj}$ of the full metric perturbation that contain additionally the rest of the emission process. 

The evolution equation for the geodesic deviation for two test particles with the tangent vector $\dot \gamma$ with initial distance $(0,ds,0,0)$ in the $x$-direction derives from equation (\ref{eq:geodesicdev}) as
\begin{equation}
\frac{D^2s^x}{d\lambda^2}=ds((R_{x00x}\dot \gamma^0+2R_{x0yx}\dot \gamma^y + 2R_{x0zx}\dot \gamma^z)\dot \gamma^0+2R_{xyzx}\dot \gamma^y \dot \gamma^z + R_{xyyx}(\dot \gamma^y)^2 + R_{xzzx}(\dot \gamma^z)^2)\,.
\end{equation}
For all $\dot \gamma$, this is dominated by the transversal curvature components due to the pulse alone if $\rho\ll r$. For infinitesimally neighbored geodesics with parallel tangent vector, we have in first order in the metric perturbation for $s=(0,ds,0,0)$
\begin{equation}
	\frac{D^2s^x}{d\lambda^2}=\ddot s^x + \dot\gamma^\rho\dot \gamma^\mu s^\nu \partial_\rho{\Gamma^{x}}_{\mu\nu}=\ddot s^x+\frac{1}{2}\dot\gamma^\rho\dot \gamma^\mu ds \partial_\rho\partial_\mu\left( h^\mathrm{e}_{xx}+ h^\mathrm{a}_{xx}\right)\,
\end{equation}
where, for $\rho\ll r$, the second term can be assumed to be much smaller than $\frac{D^2s^x}{d\lambda^2}$ since the latter is dominated by the transversal curvature components that go like $1/\rho^2$ in that limit.

\section*{References}

\bibliographystyle{ieeetr} 
\bibliography{gravphoton}

\end{document}